\documentclass{basi}
\usepackage{basi-fix}
\usepackage{txfonts}
\usepackage{graphicx}
\usepackage[authoryear]{natbib}
\bibpunct{(}{)}{;}{a}{}{,}
\usepackage{dcolumn}
\newcolumntype{d}{D{.}{.}{-1}}

\def\fdeg{^\circ\mkern-7mu.\mkern1mu}
\def\itSigma{{\mathit\Sigma}}
\def\sigmaunit{W m$^{-2}$\,Hz$^{-1}$\,sr$^{-1}$}
\def\SNR(#1.#2)#3(#4.#5){{G#1${\cdot}$#2$#3$#4${\cdot}$#5}}
\def\HI{{H\,{\sc i}}}
\def\HII{{H\,{\sc ii}}}
\def\slackbadness{\sfcode`,=3000\tolerance=3000\hbadness=3000\hfuzz=2pt}
\def\normalbadness{\sfcode`,=1250\tolerance=500\hbadness=500\hfuzz=0.5pt}
\begin{document}

\title[Galactic SNR Catalogue]{A revised Galactic supernova remnant catalogue}
\author[D.~A.\ Green]{D.~A.\ Green\thanks{e-mail:
       {\tt D.A.Green@mrao.cam.ac.uk}}\\
       Astrophysics Group, Cavendish Laboratory,
       19 J.~J.~Thomson Avenue, Cambridge CB3 0HE,\\
       United Kingdom}

\pubyear{2009}
\volume{00}
\pagerange{\pageref{firstpage}--\pageref{lastpage}}

\setcounter{page}{1}

\date{Received 2009 March} 

\maketitle

\label{firstpage}

\begin{abstract}
A revised catalogue of 274 Galactic supernova remnants (SNRs) is presented,
along with some simple statistics of their parameters. It is shown that the
remnants that have recently been identified are generally faint, as is expected
from the selection effects that apply to the identification of remnants.
\end{abstract}


\section{Introduction}\label{s:intro}

Over the last twenty five years I have produced several published versions of a
catalogue of Galactic SNRs \citep{1984MNRAS.209..449G, 1988Ap&SS.148....3G,
1991PASP..103..209G, 2002hsr..book.....S, 2004BASI...32..335G}, along with more
detailed web-based versions (most recently in 2006). Here I present an updated
version of the catalogue, now containing 274 remnants. Details of the catalogue
are presented in Section~\ref{s:catalogue}, including notes on the SNRs added
and removed since 2004. Section~\ref{s:discussion} briefly discusses some
simple statistics of the objects in the current catalogue, and reviews the
selection effects that apply to the identification of SNRs.

\section{The SNR Catalogue}\label{s:catalogue}

The current version of the catalogue contains 274 SNRs, and is based on
research in the published literature up to the end of 2008. For each remnant in
the catalogue the following parameters are given.
\begin{itemize}
\item {\bf Galactic Coordinates} of the source centroid, quoted to a tenth of a
degree as is conventional. (Note: in this catalogue additional leading zeros
are not used.)
\item {\bf Right Ascension} and {\bf Declination} of the source centroid. The
accuracy of the quoted values depends on the size of the remnant, for small
remnants they are to the nearest few seconds of time and the nearest minute of
arc respectively, whereas for larger remnants they are rounded to coarser
values, but are in every case sufficient to specify a point within the boundary
of the remnant. These coordinates are almost always deduced from radio images
rather than from X-ray or optical observations, and are for J2000.0.
\item {\bf Angular Size} of the remnant, in arcminutes, usually taken from the
highest resolution radio image available. The boundary of most remnants
approximates reasonably well to a circle or an ellipse. A single value is
quoted for the angular size of the more nearly circular remnants, which is the
diameter of a circle with an area equal to that of the remnant. For elongated
remnants the product of two values is quoted, and these are the major and minor
axes of the remnant boundary modelled as an ellipse. In a few cases an ellipse
is not a satisfactory description of the boundary of the object (refer to the
description of the individual object given in its catalogue entry), although an
angular size is still quoted for information. For `filled-centre' remnants the
size quoted is for the largest extent of the observed radio emission, not, as
at times has been used by others, the half-width of the centrally brightened
peak.
\item {\bf Type} of the SNR: `S' or `F' if the remnant shows a `shell' or
`filled-centre' structure, or `C' if it shows `composite' (or `combination')
radio structure with a combination of shell and filled-centre characteristics;
or `S?', `F?' or `C?', respectively, if there is some uncertainty, or `?' in
several cases where an object is conventionally regarded as an SNR even though
its nature is poorly known or not well-understood. (Note: the term `composite'
has been used in a different sense, by some authors, to describe SNRs with
shell radio and centrally-brightened X-ray morphologies. An alternative term
used to describe such remnants is `mixed morphology', see
\citealt{1998ApJ...503L.167R}.)
\item {\bf Flux Density} of the remnant at 1~GHz in jansky. This is {\sl not} a
measured value, but is deduced from the observed radio-frequency spectrum of
the source. The frequency of 1~GHz is chosen because flux density measurements
at frequencies both above and below this value are usually available.
\item {\bf Spectral Index} of the integrated radio emission from the remnant,
$\alpha$ (here defined in the sense, $S \propto \nu^{-\alpha}$, where $S$ is
the flux density at a frequency $\nu$), either a value that is quoted in the
literature, or one deduced from the available integrated flux densities of the
remnant. For several SNRs a simple power law is not adequate to describe their
radio spectra, either because there is evidence that the integrated spectrum is
curved or the spectral index varies across the face of the remnant. In these
cases the spectral index is given as `varies' (refer to the description of the
remnant and appropriate references in the detailed catalogue entry for more
information). In some cases, for example where the remnant is highly confused
with thermal emission, the spectral index is given as `?' since no value can be
deduced with any confidence.
\item {\bf Other Names} that are commonly used for the remnant. These are given
in parentheses if the remnant is only a part of the source. For some remnants,
notably the Crab nebula, not all common names are given.
\end{itemize}
A summary of the data available for all 274 remnants in the catalogue is given
in Table~1.

A more detailed version of the catalogue is available on the World-Wide-Web
from:
 \\[6pt]
 \centerline{\tt http://www.mrao.cam.ac.uk/surveys/snrs/}
 \\[6pt]
In addition to the basic parameters which are given in Table~1, the detailed
catalogue contains the following information.
(i) Notes if other Galactic coordinates have at times been used to label it
(usually before good observations have revealed the full extent of the object,
but sometimes in error), if the SNR is thought to be the remnant of a
historical SN, or if the nature of the source as an SNR has been questioned (in
which case an appropriate reference is usually given later in the entry).
(ii) Short descriptions of the observed structure of the remnant at radio,
X-ray and optical wavelengths, as applicable.
(iii) Notes on distance determinations, and any point sources or pulsars in or
near the object (although they may not necessarily be related to the remnant).
(iv) References to observations are given for each remnant, complete with
journal, volume, page, and a short description of what information each paper
contains (for radio observations these include the telescopes used, the
observing frequencies and resolutions, together with any flux density
determinations). These references are {\sl not} complete, but cover
representative and recent observations of the remnant -- up to the end of 2008
-- and they should themselves include references to earlier work.

The detailed version is available as postscript or pdf for downloading and
printing, or as HTML web pages for each individual remnant. The web pages
include links to the `NASA Astrophysics Data System' for each of the over two
thousand references that are included in the detailed listings for individual
SNRs.

Some of the parameters included in the catalogue are themselves of quite
variable quality. For example, the radio flux density of each remnant at 1~GHz.
This is generally of good quality, being obtained from several radio
observations over a range of frequencies, both above and below 1~GHz. However,
for a small number of remnants (16 remnants in the current catalogue) -- often
those which have been identified at other than radio wavelengths -- no reliable
radio flux density, or only a limit, is available. Also, although the detailed
version of the catalogue contains notes on distances for many remnants reported
in the literature, these have a range of reliability. Consequently the
distances given within the detailed catalogue should be used with caution in
any statistical studies.

\subsection{Changes from the 2004 version}\label{s:new}

The follow remnants have been added to the catalogue since the last published
version \citep{1984MNRAS.209..449G}.

\begin{itemize}
\item G32.4$+$0.1, identified by \cite{2004PASJ...56.1059Y}.
\item Two 2nd quadrant remnants, G96.0$+$2.0 and G113.0$+$0.2, identified by
\cite{2005A&A...444..871K}.
\item G337.2$+$0.1, which was confirmed as a SNR by \cite{2005A&A...431L...9C}.
\slackbadness
\item 31 new SNRs in the region $4\fdeg5 <l < 22\fdeg0$, $|b| <1\fdeg25$
(G5.5$+$0.3, G6.1$+$0.5, G6.5$-$0.4, G7.2$+$0.2, G8.3$-$0.0, G8.9$+$0.4,
G9.7$-$0.0, G9.9$-$0.8, G10.5$-$0.0, G11.0$-$0.0, G11.1$-$0.7, G11.1$-$1.0,
G11.1$+$0.1, G11.8$-$0.2, G12.2$+$0.3, G12.5$+$0.2, G12.7$-$0.0, G12.8$-$0.0,
G14.1$-$0.1, G14.3$+$0.1, G15.4$+$0.1, G16.0$-$0.5, G16.4$-$0.5, G17.0$-$0.0,
G17.4$-$0.1, G18.1$-$0.1, G18.6$-$0.2, G19.1$+$0.2, G20.4$+$0.1, G21.0$-$0.4
and G21.5$-$0.1) identified by \cite{2006ApJ...639L..25B}. (There are the 31
objects classed as `I' or `II', i.e.\ those thought to be very or fairly
confidently identified as SNRs by \citeauthor{2006ApJ...639L..25B})
\normalbadness
\item G83.0$-$0.3, which had been suggested as a SNR by
\cite{1992AJ....103..931T}, and is now included in the catalogue following
improved observations by \cite{2006A&A...457.1081K} which confirm its nature.
\item G108.2$-$0.6, identified by \cite{2007A&A...465..907T}.
\item G315.1$+$2.7, which had been suggested as a candidate SNR by
\cite{1995MNRAS.277...36D, 1997MNRAS.287..722D}, and was confirmed as a SNR by
optical and radio survey observations by \cite{2007MNRAS.374.1441S}.
\item G327.2$-$0.1, a shell remnant found around the magnetar 1E 1547.0$-$5408,
see \cite{2007ApJ...667.1111G}.
\item G332.5$-$5.6, identified by \cite{2007MNRAS.375...92R}.
\item G350.1$-$0.3, which was listed in early versions of the catalogue, but
was removed from the version in \cite{1991PASP..103..209G}, as observations by
\cite{1986A&A...162..217S} did not allow a clear identification of the nature
of this source. Recently \cite{2008ApJ...680L..37G} have presented new
observations of this source, including \HI\ absorption observations which
indicate it is Galactic, which -- along with other observations, including its
X-ray emission -- support an SNR identification. However, its structure at
radio wavelengths is rather different from other known remnants.
\item G353.6$-$0.7, a shell remnant associated with HESS J1731$-$347 identified
by \cite{2008ApJ...679L..85T}.
\item Three sources -- G355.4$+$0.7, G358.1$+$1.0, G358.5$-$0.9 -- which had
been identified as possible SNRs by \cite{1994MNRAS.270..847G}, have now been
added to the catalogue, following further observations by
\cite{2006JPhCS..54..152R} which confirm their nature.
\end{itemize}
Since 2004 two objects have been removed from the catalogue, as they have been
identified as \HII\ regions, namely: G166.2$+$2.5 ($=$OA 184), see
\cite{2006A&A...454..517F}; and G84.9$+$0.5, see \cite{2007ApJ...667..248F},
and also see \cite{2006A&A...457.1081K}.

As noted above, the detailed catalogue is based on published papers up to the
end of 2008, and hence the catalogue does not include remnants identified more
recently (e.g.\ \citealt{2009arXiv0904.3657G, 2009arXiv0903.4317H}).

\subsection{Possible and Probable SNRs}

In addition to the observational selection effects that are discussed further
in Section~\ref{s:selection}, it should be noted that the catalogue is far from
homogeneous. It is particularly difficult to be uniform in terms of which
objects are considered as definite remnants, and are included in the catalogue,
rather than listed as possible or probable remnants which require further
observations to clarify their nature. Although many remnants, or possible
remnants, were first identified from wide-area radio surveys, many others have
been observed with diverse observational parameters, making uniform criteria
for inclusion in the main catalogue difficult. The detailed version of the
catalogue contains notes both on those objects no longer thought to be SNRs,
and on many possible and probable remnants that have been reported in the
literature.

\begin{figure}
\centerline{\includegraphics[width=11.0cm]{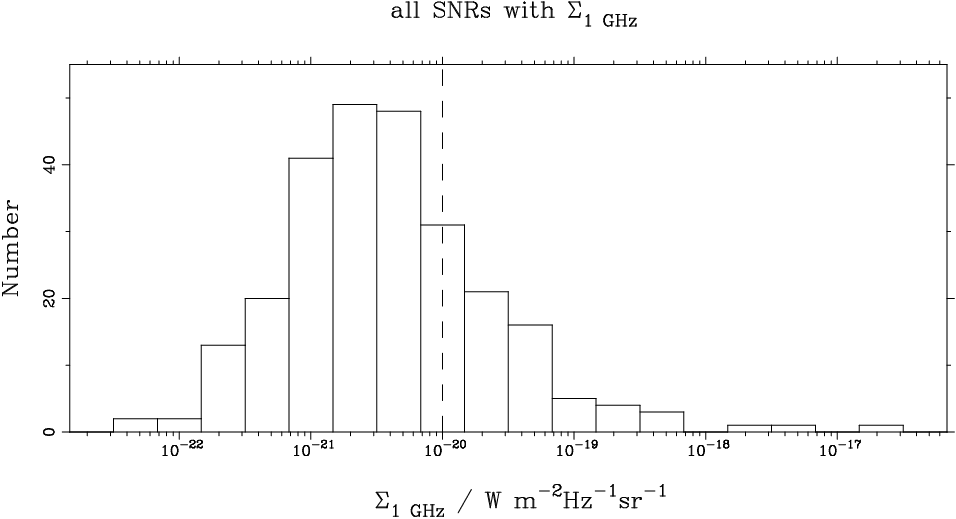}}
\bigskip
\centerline{\includegraphics[width=11.0cm]{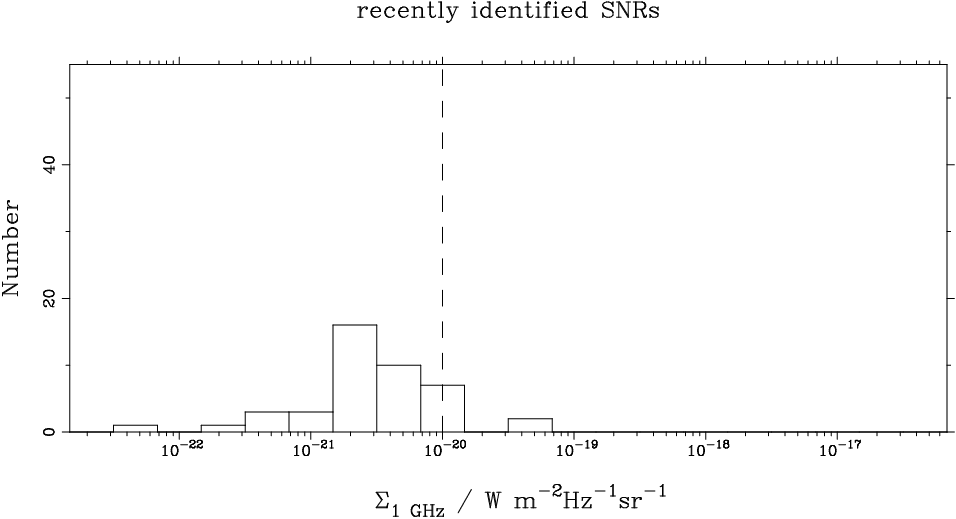}}
\caption{Histograms in surface brightness at 1~GHz for (top) all 258 Galactic
SNRs with 1-GHz flux densities, and (bottom) the remnants included in the
catalogue since 2004. The dashed lines mark the nominal completeness limit of
the catalogue, see text.\label{f:sigma}}
\end{figure}

\begin{figure}
\centerline{\includegraphics[width=11.0cm]{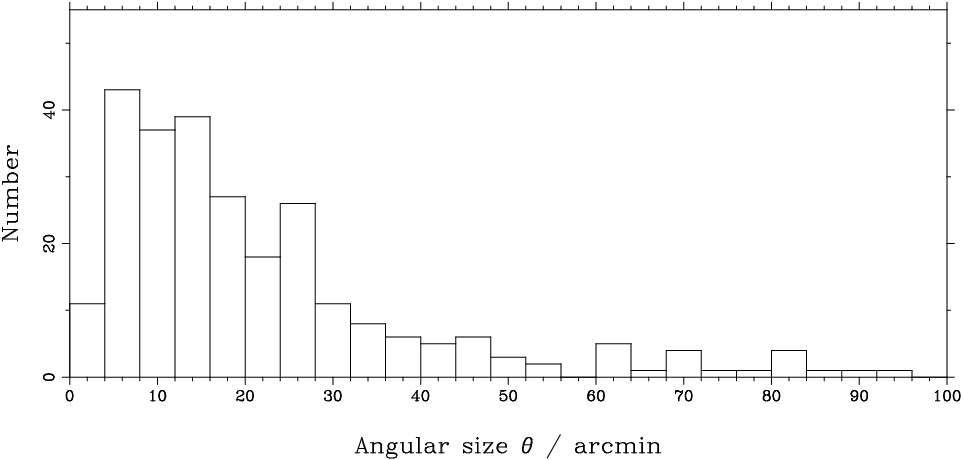}}
\caption{Histogram of the angular size of 261 Galactic SNRs (13 remnants larger
than 100 arcmin are not included).\label{f:theta}}
\end{figure}

\begin{figure}
\centerline{\includegraphics[width=13.0cm]{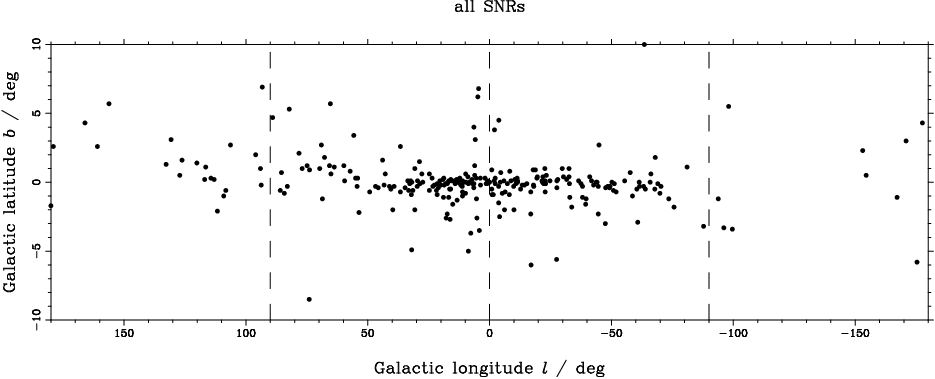}}
\bigskip
\centerline{\includegraphics[width=13.0cm]{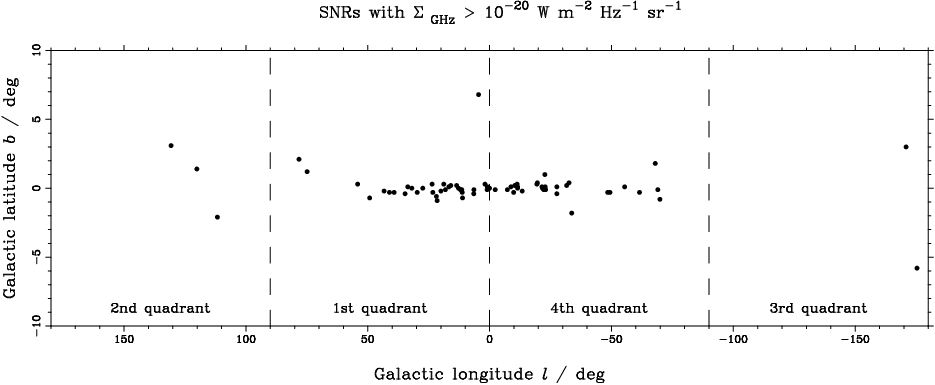}}
\caption{Galactic distribution of (top) all Galactic SNR and (bottom) those
SNRs with a surface brightness at 1~GHz greater than $10^{-20}$ {\sigmaunit}.
(Note that the latitude and longitude axes are not on the same
scale.)\label{f:lb}}
\end{figure}

\section{Discussion}\label{s:discussion}

\subsection{Some simple statistics}

There are 16 Galactic SNRs that are either not detected at radio wavelengths,
or are poorly defined by current radio observations, so that their flux density
at 1~GHz cannot be determined with any confidence: i.e.\ 94\% have a flux
density at 1~GHz included in the catalogue. Of the catalogued remnants,
$\approx 40$\% are detected in X-ray, and $\approx 20$\% in the optical. At
both of these wavebands, Galactic absorption hampers the detection of distant
remnants.

In the current version of the catalogue, 78\% of remnants are classed as shell
(or possible shell), 12\% are composite (or possible composite), and 4\% are
filled-centre (or possible filled centre) remnants. The type of the remaining
remnants is not clear from current observations, or else they are objects which
are conventionally regarded as SNRs although they do not fit well into any of
the conventional types (e.g.\ CTB80 ($=$\SNR(69.0)+(2.7)), MSH 17$-$3{\em 9}
($=$\SNR(357.7)-(0.1))).

\subsection{Selection effects}\label{s:selection}

Although several Galactic SNRs have been identified at other than radio
wavelengths, in practice the dominant selection effects are those that are
applicable at radio wavelengths. Simplistically, two selection effects apply to
the identification of Galactic SNRs due to the difficulty in identifying (i)
faint remnants and (ii) small angular size remnants (see
\citealt{1991PASP..103..209G, 2004BASI...32..335G, 2005MmSAI..76..534G} for
more detailed discussions).

In terms of surface brightness, which is distance-independent, the wide-area
radio surveys covering the Galactic plane mean that the catalogue is thought to
be complete down to about $\itSigma_{\rm 1~GHz} \approx 10^{-20}$ {\sigmaunit}
\citep{2004BASI...32..335G, 2005MmSAI..76..534G}. Fig.~\ref{f:sigma} shows a
histogram of $\itSigma_{\rm 1~GHz}$ for the 258 remnants in the current
catalogue for which 1-GHz flux densities are available, and also for those
remnants included in the catalogue since the 2004 version. This shows that the
majority of newly identified remnants are indeed fainter than nominal
completeness limit of $\approx 10^{-20}$ {\sigmaunit}. Most of these are from
the deep survey of a limited region of the Galactic plane by
\cite{2006ApJ...639L..25B} (see Section~\ref{s:new} above). The are a few
newly-identified remnants that are brighter than the nominal surface brightness
completeness limit of the catalogue of $\approx 10^{-20}$ {\sigmaunit}. The
brightest two of these, namely \SNR(337.2)+(0.1) and \SNR(350.1)-(0.3), both
have relatively small angular sizes ($\le 4$ arcmin). Such small SNRs are
difficult to identify because their structure is not easily recognised without
high enough resolution observations. It is difficult to quantify the angular
size selection effect -- see \cite{2004BASI...32..335G} for some further
discussion -- but Fig.~\ref{f:theta} shows a histogram of the observed angular
sizes of remnants in the current catalogue. (Note: for elongated remnants,
which have angular sizes given as  $n \times m$ arcmin$^2$ in the catalogue, a
single diameter of $\sqrt{nm}$ has been used in this histogram.) Since much of
the Galactic plane is now covered by several recent radio surveys -- the
Canadian/VLA/Southern Galactic Plane Surveys, see \cite{1998PASA...15...56E,
2001ApJ...551..394M, 2003AJ....125.3145T, 2006AJ....132.1158S} -- with a
resolution of $\approx 1$~arcmin, systematic searches using these should be
able to identify small angular size remnants that have not been recognised in
previous surveys with lower resolution.

Due to these selection effects, care has to be taken in using the catalogue for
statistical studies. For example, Fig.~\ref{f:lb} shows the distribution of
both (a) {\sl all} SNRs, and (b) the brighter ones (with $\itSigma_{\rm 1~GHz}
\ge 10^{-20}$ {\sigmaunit}), to illustrate the surface brightness selection
effect. This shows that relatively many more remnants are seen in the 2nd and
3rd Galactic quadrants and away from $b=0^\circ$, when looking at the whole
catalogue rather than the brighter remnants. This is expected, as the Galactic
background is fainter in these regions, and hence it is easier to identify
faint SNRs here.

In addition to selection effects that apply to the identification of Galactic
SNRs, many statistical studies also require distances for all remnants. The
surface brightness--diameter (`$\itSigma{-}D$') relation is often used for such
studies, but -- as has been discussed in \cite{2004BASI...32..335G,
2005MmSAI..76..534G} -- there are problems with this method given the large
range of properties shown by Galactic SNRs, and the observational selection
effects. Also, as discussed in detail in \cite{2005MmSAI..76..534G}, it should
be noted that some $\itSigma{-}D$ relations in the literature (e.g.\
\citealt{1998ApJ...504..761C,2007Ap&SS.307..423S}) have been derived using
least-squares regression minimising deviation in terms of $\log(\itSigma)$,
rather than in terms of $\log(D)$. As the $\itSigma{-}D$ relation is used to
derive a diameter (and hence distance) from an observed surface brightness,
then least-squares in terms of $\log(D)$ is appropriate. Given the large
scatter in the observed properties of SNRs with known distances, then using a
least-square regression in terms of $\log(\itSigma)$ leads to a significantly
flatter $\itSigma{-}D$ slope than if a regression in terms of $\log(D)$ is
used. This leads to an overestimate of the diameters, and hence distances, for
all faint remnants.

\section*{Acknowledgements}

I am grateful to many colleagues for numerous comments on, and corrections to,
the various versions of the Galactic SNR catalogue. This research has made use
of NASA's Astrophysics Data System Bibliographic Services.

\setlength{\bibsep}{0pt}
\def\newblock{}

\newpage
\clearpage
\newcount\linesdone
\global\linesdone=0
\newcount\processed
\global\processed=0
%
%
\def\captiontext{Galactic Supernova Remnants: summary data.}
\def\tops{%
  \setbox0=\vbox\bgroup%
  \centerline{\small{\bf Table~1}. \captiontext}
  \centerline{\hrulefill}
  \smallskip
  \centerline{%
    \hbox to 0.06\hsize{\hfil$l$\enskip}%
    \hbox to 0.065\hsize{\hfil$b$\enskip}%
    \hbox to 0.195\hsize{\hss\quad RA (J2000) Dec\hss}%
    \hbox to 0.12\hsize{\hfil size\hfil}%
    \hbox to 0.05\hsize{type\hfil}%
    \hbox to 0.085\hsize{\hfil Flux at\hfil}%
    \hbox to 0.10\hsize{\hfil spectral\hfil}%
    \hbox to 0.329\hsize{\enspace other \hfil}%
    \hfill
  }
  \centerline{%
    \hbox to 0.125\hsize{\hfil}%
    \hbox to 0.11\hsize{\hfil$({\rm h}$\enskip${\rm m}$\enskip${\rm s})$}%
    \hbox to 0.085\hsize{\hfil$({}^\circ$\kern9pt${}'$)}%
    \hbox to 0.12\hsize{\hfil /arcmin\hfil}%
    \hbox to 0.05\hsize{}%
    \hbox to 0.085\hsize{\hss 1~GHz/Jy\hss}%
    \hbox to 0.10\hsize{\hfil index\hfil}%
    \hbox to 0.329\hsize{\enspace name(s)\hfil}%
    \hfill
  }
  \centerline{\hrulefill} %
  \medskip
  \egroup\box0
}
%
%
%
%
%
%
%
%
%
%
%
%
\def\LONGITUDE #1 {\def\ldegrees{#1}}
\def\LATITUDE #1 {\def\bdegrees{#1}}
\def\RAHMS #1 #2 #3 {\def\rahms{#1~#2~#3}}
\def\DECDM #1 #2 {\def\decdm{#1~#2}}
\def\SIZE #1 {\def\size{#1}}
\def\ALPHA #1 {\def\spectralindex{#1}}
\def\FLUX1GHZ #1 {\def\flux1GHz{#1}}
\def\TYPE #1 {\def\type{#1}}
\def\NAMES #1\par{\def\names{#1}\ifnum\linesdone=0\tops\fi%
  \global\advance\linesdone by 1
  \global\advance\processed by 1
  \centerline{%
    \hbox to 0.06\hsize{\hfil$\ldegrees$}%
    \hbox to 0.065\hsize{\hfil$\bdegrees$}%
    \hbox to 0.11\hsize{\hfil$\rahms$}%
    \hbox to 0.085\hsize{\hfil$\decdm$}%
    \hbox to 0.12\hsize{\hfil$\size$\hfil}%
    \hbox to 0.05\hsize{\enspace\type\hfil}%
    \hbox to 0.085\hsize{\hss\flux1GHz\hss}%
    \hbox to 0.10\hsize{\quad\spectralindex\hfil}%
    \hbox to 0.325\hsize{\enspace\names\hfil}%
    \hfill
  }
  \setbox0=\vbox\bgroup \eightpoint\hfuzz=20pt
}
%
%
\def\NOTES{\par}
\def\RADIO{\par}
\def\XRAY{\par}
\def\OPTICAL{\par}
\def\DISTANCE{\par}
\def\POINT{\par}
\def\REFERENCE{\par}
%
%
\def\PM/{\pm }
\def\X/{\times }
\def\I/{{\small I}}
\def\II/{{\small II}}
\def\III/{{\small III}}
\def\V/{{\small V}}
\def\VV/{{\small X}}       
\def\XIV/{{\small XIV}}
\def\AD/{{\small AD}}
\def\Halpha/{{H$\alpha$}}
\def\HCOplus/{{HCO$^+$}}
\def\etal/{{et al.}}
\def\fdeg{^\circ\mkern-7mu.\mkern1mu}
\def\fmin{'\mkern-5mu.}
\def\fsec{''\mkern-7mu.\mkern1mu}
\let\eightpoint=\footnotesize
%
%
\def\DATE #1\par{
  \egroup
  \ifnum\linesdone=35
    \global\linesdone=0
    \global\processed=0
    \centerline{\hrulefill}
    \vfill\eject
    \def\captiontext{(continued).}
  \fi
  \ifnum\processed=5
    \vskip 6pt plus 2pt minus 1pt
    \global\processed=0
  \fi
}
%
%
\LONGITUDE 0.0 \LATITUDE +0.0
\RAHMS 17 45 44 \DECDM -29 00
\SIZE 3.5\X/2.5 \TYPE S
\FLUX1GHZ 100? \ALPHA 0.8?
\NAMES Sgr A East

\RADIO Non-thermal shell, in complex region, interacting with molecular material to the west.
\XRAY Diffuse emission, centrally peaked.
\POINT Compact X-ray source.
\DATE 18 Feb 2009

\LONGITUDE 0.3 \LATITUDE +0.0
\RAHMS 17 46 15 \DECDM -28 38
\SIZE 15\X/8 \TYPE S
\FLUX1GHZ 22 \ALPHA 0.6
\NAMES

\NOTES Has been called G0.33$+$0.04 and G0.4$+$0.1.
\RADIO Bilateral shell, near Galactic Centre.
\DATE 31 Mar 2006

\LONGITUDE 0.9 \LATITUDE +0.1
\RAHMS 17 47 21 \DECDM -28 09
\SIZE 8 \TYPE C
\FLUX1GHZ 18? \ALPHA varies
\NAMES

\RADIO Flat spectrum core within steep spectrum shell.
\XRAY Central core, with non-thermal spectrum.
\DATE 18 Feb 2009

\LONGITUDE 1.0 \LATITUDE -0.1
\RAHMS 17 48 30 \DECDM -28 09
\SIZE 8 \TYPE S
\FLUX1GHZ 15 \ALPHA 0.6?
\NAMES

\NOTES Has been called G1.05$-$0.1 and G1.05$-$0.15.
\RADIO Incomplete shell, to the S of Sgr~D.
\XRAY Possibly detected.
\DATE 18 Feb 2009

\LONGITUDE 1.4 \LATITUDE -0.1
\RAHMS 17 49 39 \DECDM -27 46
\SIZE 10 \TYPE S
\FLUX1GHZ 2? \ALPHA ?
\NAMES

\RADIO Shell, brightest in E.
\DATE 18 Feb 2009

\LONGITUDE 1.9 \LATITUDE +0.3
\RAHMS 17 48 45 \DECDM -27 10
\SIZE 1.5 \TYPE S
\FLUX1GHZ 0.6 \ALPHA 0.6
\NAMES

\RADIO Shell, brighter to the N, brightening.
\XRAY Shell, with bright limbs to E and W.
\DATE 18 Feb 2009

\LONGITUDE 3.7 \LATITUDE -0.2
\RAHMS 17 55 26 \DECDM -25 50
\SIZE 14\X/11 \TYPE S
\FLUX1GHZ 2.3 \ALPHA 0.65
\NAMES

\NOTES Has been called G003.8$-$00.3.
\RADIO Double arc.
\DATE 31 Mar 2006

\LONGITUDE 3.8 \LATITUDE +0.3
\RAHMS 17 52 55 \DECDM -25 28
\SIZE 18 \TYPE S?
\FLUX1GHZ 3? \ALPHA 0.6
\NAMES

\RADIO Incomplete shell.
\DATE 11 Jan 2004

\LONGITUDE 4.2 \LATITUDE -3.5
\RAHMS 18 08 55 \DECDM -27 03
\SIZE 28 \TYPE S
\FLUX1GHZ 3.2? \ALPHA 0.6?
\NAMES

\RADIO Elongated shell.
\DATE 19th May 1992

\LONGITUDE 4.5 \LATITUDE +6.8
\RAHMS 17 30 42 \DECDM -21 29
\SIZE 3 \TYPE S
\FLUX1GHZ 19 \ALPHA 0.64
\NAMES Kepler, SN1604, 3C358

\NOTES This is the remnant of Kepler's SN of \AD/1604.
\RADIO Incomplete shell, brighter to the N.
\OPTICAL Faint filaments.
\XRAY Shell, brighter to the N.
\DISTANCE Optical expansion and proper motion indicates about 2.9~kpc, H\I/ observations suggest 3.4 to 6.4~kpc.
\DATE 18 Feb 2009

\LONGITUDE 4.8 \LATITUDE +6.2
\RAHMS 17 33 25 \DECDM -21 34
\SIZE 18 \TYPE S
\FLUX1GHZ 3 \ALPHA 0.6
\NAMES

\NOTES Has been called G4.5$+$6.2.
\RADIO Faint shell.
\DATE 12 Oct 2001

\LONGITUDE 5.2 \LATITUDE -2.6
\RAHMS 18 07 30 \DECDM -25 45
\SIZE 18 \TYPE S
\FLUX1GHZ 2.6? \ALPHA 0.6?
\NAMES

\RADIO Poorly resolved shell.
\DATE 19th May 1992

\LONGITUDE 5.4 \LATITUDE -1.2
\RAHMS 18 02 10 \DECDM -24 54
\SIZE 35 \TYPE C?
\FLUX1GHZ 35? \ALPHA 0.2?
\NAMES Milne 56

\NOTES Part been called G5.3$-$1.0. Has been suggested that this is not a SNR.
\RADIO Incomplete shell, including wide `v' of emission to east with small flat-spectrum source at apex.
\OPTICAL Detected.
\XRAY Pulsar detected, with faint extension.
\DISTANCE H\I/ absorption suggests $>4.3$~kpc.
\POINT Pulsar nearby, in flat spectrum source.
\DATE 18 Feb 2009

\LONGITUDE 5.5 \LATITUDE +0.3
\RAHMS 17 57 04 \DECDM -24 00
\SIZE 15\X/12 \TYPE S
\FLUX1GHZ 5.5 \ALPHA 0.7
\NAMES

\NOTES Has been called G5.55$+$0.32.
\RADIO Shell.
\DATE 2006 March 29

\LONGITUDE 5.9 \LATITUDE +3.1
\RAHMS 17 47 20 \DECDM -22 16
\SIZE 20 \TYPE S
\FLUX1GHZ 3.3? \ALPHA 0.4?
\NAMES

\RADIO Asymmetric shell.
\DATE 19th May 1992

\LONGITUDE 6.1 \LATITUDE +0.5
\RAHMS 17 57 29 \DECDM -23 25
\SIZE 18\X/12 \TYPE S
\FLUX1GHZ 4.5 \ALPHA 0.9
\NAMES

\NOTES Has been called G6.10$+$0.53.
\RADIO Partial shell.
\DATE 2006 March 29

\LONGITUDE 6.1 \LATITUDE +1.2
\RAHMS 17 54 55 \DECDM -23 05
\SIZE 30\X/26 \TYPE F
\FLUX1GHZ 4.0? \ALPHA 0.3?
\NAMES

\NOTES Has been called G6.1$+$1.15.
\RADIO Faint, diffuse emission.
\DATE 28 Jun 1995

\LONGITUDE 6.4 \LATITUDE -0.1
\RAHMS 18 00 30 \DECDM -23 26
\SIZE 48 \TYPE C
\FLUX1GHZ 310 \ALPHA varies
\NAMES W28

\NOTES Has been called G6.6$-$0.2.
\RADIO Several non-thermal sources in a ring, with flat spectrum core.
\OPTICAL Diffuse emission.
\XRAY Diffuse emission from most of the remnant.
\POINT Young pulsar near edge of remnant, but not thought to be related.
\DISTANCE H\I/ observations suggest 1.9~kpc.
\DATE 18 Feb 2009

\LONGITUDE 6.4 \LATITUDE +4.0
\RAHMS 17 45 10 \DECDM -21 22
\SIZE 31 \TYPE S
\FLUX1GHZ 1.3? \ALPHA 0.4?
\NAMES

\RADIO Faint asymmetric shell.
\DATE 19th May 1992

\LONGITUDE 6.5 \LATITUDE -0.4
\RAHMS 18 02 11 \DECDM -23 34
\SIZE 18 \TYPE S
\FLUX1GHZ 27 \ALPHA 0.6
\NAMES

\NOTES Has been called G6.51$-$0.48, and part has been called G6.67$-$0.42.
\RADIO Shell.
\DATE 18 Feb 2009

\LONGITUDE 7.0 \LATITUDE -0.1
\RAHMS 18 01 50 \DECDM -22 54
\SIZE 15 \TYPE S
\FLUX1GHZ 2.5? \ALPHA 0.5?
\NAMES

\NOTES Has been called G7.06$-$0.12.
\RADIO Double rim, brightest in W, confused by bright H\II/ region M20 in SE.
\DATE 10th October 2001

\LONGITUDE 7.2 \LATITUDE +0.2
\RAHMS 18 01 07 \DECDM -22 38
\SIZE 12 \TYPE S
\FLUX1GHZ 2.8 \ALPHA 0.6
\NAMES

\NOTES Has been called G7.20$+$0.20.
\RADIO Partial shell.
\DATE 2006 March 29

\LONGITUDE 7.7 \LATITUDE -3.7
\RAHMS 18 17 25 \DECDM -24 04
\SIZE 22 \TYPE S
\FLUX1GHZ 11 \ALPHA 0.32
\NAMES 1814$-$24

\RADIO Shell, with high polarisation.
\DATE 20 Aug 1996

\LONGITUDE 8.3 \LATITUDE -0.0
\RAHMS 18 04 34 \DECDM -21 49
\SIZE 5\X/4 \TYPE S
\FLUX1GHZ 1.2 \ALPHA 0.6
\NAMES

\NOTES Has been called G8.31$-$0.09.
\RADIO Shell.
\DATE 18 Feb 2009

\LONGITUDE 8.7 \LATITUDE -5.0
\RAHMS 18 24 10 \DECDM -23 48
\SIZE 26 \TYPE S
\FLUX1GHZ 4.4 \ALPHA 0.3
\NAMES

\RADIO Asymmetric shell.
\DATE 19th May 1992

\LONGITUDE 8.7 \LATITUDE -0.1
\RAHMS 18 05 30 \DECDM -21 26
\SIZE 45 \TYPE S?
\FLUX1GHZ 80 \ALPHA 0.5
\NAMES (W30)

\NOTES Has been called G8.6$-$0.1.
\RADIO Clumpy non-thermal shell, with low-frequency turnover.
\XRAY Northern edge detected.
\POINT Pulsar inside western edge.
\DATE 16 Mar 2009

\LONGITUDE 8.9 \LATITUDE +0.4
\RAHMS 18 03 58 \DECDM -21 03
\SIZE 24 \TYPE S
\FLUX1GHZ 9 \ALPHA 0.6
\NAMES

\NOTES Has been called G8.90$+$0.40.
\RADIO Shell.
\DATE 2006 March 29

\LONGITUDE 9.7 \LATITUDE -0.0
\RAHMS 18 07 22 \DECDM -20 35
\SIZE 15\X/11 \TYPE S
\FLUX1GHZ 3.7 \ALPHA 0.6
\NAMES

\NOTES Has been called G9.7$-$0.1 and G9.70$-$0.06.
\RADIO Shell.
\DATE 2006 March 29

\LONGITUDE 9.8 \LATITUDE +0.6
\RAHMS 18 05 08 \DECDM -20 14
\SIZE 12 \TYPE S
\FLUX1GHZ 3.9 \ALPHA 0.5
\NAMES

\RADIO Asymmetric shell.
\DATE 13 Aug 1998

\LONGITUDE 9.9 \LATITUDE -0.8
\RAHMS 18 10 41 \DECDM -20 43
\SIZE 12 \TYPE S
\FLUX1GHZ 6.7 \ALPHA 0.4
\NAMES

\NOTES Has been called G9.95$-$0.81.
\RADIO Shell.
\DATE 2006 March 29

\LONGITUDE 10.5 \LATITUDE -0.0
\RAHMS 18 09 08 \DECDM -19 47
\SIZE 6 \TYPE S
\FLUX1GHZ 0.9 \ALPHA 0.6
\NAMES

\NOTES Has been called G10.59$-$0.04.
\RADIO Partial shell.
\XRAY Detected.
\DATE 2006 March 29

\LONGITUDE 11.0 \LATITUDE -0.0
\RAHMS 18 10 04 \DECDM -19 25
\SIZE 11\X/9 \TYPE S
\FLUX1GHZ 1.3 \ALPHA 0.6
\NAMES

\NOTES Has been called G11.0$+$0.0 and G11.03$-$0.05.
\RADIO Partial shell.
\XRAY Diffuse emission.
\DATE 2006 March 29

\LONGITUDE 11.1 \LATITUDE -1.0
\RAHMS 18 14 03 \DECDM -19 46
\SIZE 18\X/12 \TYPE S
\FLUX1GHZ 5.8 \ALPHA 0.6
\NAMES

\NOTES Has been called G11.2$-$1.1 and G11.17$-$1.04.
\RADIO Shell.
\DATE 2006 March 29

\LONGITUDE 11.1 \LATITUDE -0.7
\RAHMS 18 12 46 \DECDM -19 38
\SIZE 11\X/7 \TYPE S
\FLUX1GHZ 1.0 \ALPHA 0.7
\NAMES

\NOTES Has been called G11.15$-$0.71.
\RADIO Partial shell.
\DATE 2006 March 29

\LONGITUDE 11.1 \LATITUDE +0.1
\RAHMS 18 09 47 \DECDM -19 12
\SIZE 12\X/10 \TYPE S
\FLUX1GHZ 2.3 \ALPHA 0.4
\NAMES

\NOTES Has been called G11.18$+$0.11.
\RADIO Shell.
\DATE 2006 March 29

\LONGITUDE 11.2 \LATITUDE -0.3
\RAHMS 18 11 27 \DECDM -19 25
\SIZE 4 \TYPE C
\FLUX1GHZ 22 \ALPHA 0.6
\NAMES

\NOTES Probably associated with the SN of \AD/386.
\RADIO Symmetrical clumpy shell, with flatter spectrum core.
\XRAY Shell, with hard spectrum centrally brightened region around pulsar.
\POINT Central pulsar.
\DISTANCE H\I/ absorption indicates 4.4~kpc.
\DATE 16 Mar 2009

\LONGITUDE 11.4 \LATITUDE -0.1
\RAHMS 18 10 47 \DECDM -19 05
\SIZE 8 \TYPE S?
\FLUX1GHZ 6 \ALPHA 0.5
\NAMES

\RADIO Incomplete shell, possibly with central core.
\DATE 18 Apr 2006

\LONGITUDE 11.8 \LATITUDE -0.2
\RAHMS 18 12 25 \DECDM -18 44
\SIZE 4 \TYPE S
\FLUX1GHZ 0.7 \ALPHA 0.3
\NAMES

\NOTES Has been called G11.89$-$0.21.
\RADIO Shell.
\XRAY Detected.
\DATE 2006 March 29

\LONGITUDE 12.0 \LATITUDE -0.1
\RAHMS 18 12 11 \DECDM -18 37
\SIZE 7? \TYPE ?
\FLUX1GHZ 3.5 \ALPHA 0.7
\NAMES

\RADIO Incomplete shell, defined in E only.
\XRAY Detected.
\DATE 10 Oct 2001

\LONGITUDE 12.2 \LATITUDE +0.3
\RAHMS 18 11 17 \DECDM -18 10
\SIZE 6\X/5 \TYPE S
\FLUX1GHZ 0.8 \ALPHA 0.7
\NAMES

\NOTES Has been called G12.26$+$0.30.
\RADIO Partial shell.
\DATE 2006 March 29

\LONGITUDE 12.5 \LATITUDE +0.2
\RAHMS 18 12 14 \DECDM -17 55
\SIZE 6\X/5 \TYPE C?
\FLUX1GHZ 0.6 \ALPHA 0.4
\NAMES

\NOTES Has been called G12.58$+$0.22.
\RADIO Diffuse, central brightened.
\DATE 2006 March 29

\LONGITUDE 12.7 \LATITUDE -0.0
\RAHMS 18 13 19 \DECDM -17 54
\SIZE 6 \TYPE S
\FLUX1GHZ 0.8 \ALPHA 0.8
\NAMES

\NOTES Has been called G12.72$-$0.00.
\RADIO Shell.
\DATE 2006 March 29

\LONGITUDE 12.8 \LATITUDE -0.0
\RAHMS 18 13 37 \DECDM -17 49
\SIZE 3 \TYPE C?
\FLUX1GHZ 0.8 \ALPHA 0.5
\NAMES

\NOTES Has been called G12.82$-$0.02 and G12.83$-$0.02.
\RADIO Shell.
\XRAY Diffuse, with central source.
\DATE 16 Mar 2009

\LONGITUDE 13.3 \LATITUDE -1.3
\RAHMS 18 19 20 \DECDM -18 00
\SIZE 70\X/40 \TYPE S?
\FLUX1GHZ ? \ALPHA ?
\NAMES

\RADIO Amorphous emission.
\XRAY Elongated emission.
\OPTICAL Filaments in S.
\DISTANCE Absorption indicates 2--4 kpc.
\DATE 1 Aug 2000

\LONGITUDE 13.5 \LATITUDE +0.2
\RAHMS 18 14 14 \DECDM -17 12
\SIZE 5\X/4 \TYPE S
\FLUX1GHZ 3.5? \ALPHA 1.0?
\NAMES

\NOTES Has been called G13.46$+$0.16.
\RADIO Elongated, incomplete shell.
\DATE 13 Aug 1998

\LONGITUDE 14.1 \LATITUDE -0.1
\RAHMS 18 15 52 \DECDM -16 34
\SIZE 6\X/5 \TYPE S
\FLUX1GHZ 0.5 \ALPHA 0.6
\NAMES

\NOTES Has been called G14.18$-$0.12.
\RADIO Shell.
\DATE 2006 March 29

\LONGITUDE 14.3 \LATITUDE +0.1
\RAHMS 18 15 58 \DECDM -16 27
\SIZE 5\X/4 \TYPE S
\FLUX1GHZ 0.6 \ALPHA 0.4
\NAMES

\NOTES Has been called G14.30$+$0.14.
\RADIO Partial shell.
\DATE 2006 March 29

\LONGITUDE 15.1 \LATITUDE -1.6
\RAHMS 18 24 00 \DECDM -16 34
\SIZE 30\X/24 \TYPE S
\FLUX1GHZ 5.5? \ALPHA 0.8?
\NAMES

\RADIO Elongated, incomplete shell.
\OPTICAL Diffuse shell.
\DATE 18 Feb 2009

\LONGITUDE 15.4 \LATITUDE +0.1
\RAHMS 18 18 02 \DECDM -15 27
\SIZE 15\X/14 \TYPE S
\FLUX1GHZ 5.6 \ALPHA 0.6
\NAMES

\NOTES Has been called G15.42$+$0.18.
\RADIO Shell.
\DATE 2006 March 29

\LONGITUDE 15.9 \LATITUDE +0.2
\RAHMS 18 18 52 \DECDM -15 02
\SIZE 7\X/5 \TYPE S?
\FLUX1GHZ 5 \ALPHA 0.6?
\NAMES

\RADIO Incomplete shell, with bright concentration to the E.
\XRAY Shell, brighter to S and E.
\DATE 18 Feb 2009

\LONGITUDE 16.0 \LATITUDE -0.5
\RAHMS 18 21 56 \DECDM -15 14
\SIZE 15\X/10 \TYPE S
\FLUX1GHZ 2.7 \ALPHA 0.6
\NAMES

\NOTES Has been called G16.05$-$0.57.
\RADIO Shell.
\DATE 2006 March 29

\LONGITUDE 16.2 \LATITUDE -2.7
\RAHMS 18 29 40 \DECDM -16 08
\SIZE 17 \TYPE S
\FLUX1GHZ 2 \ALPHA 0.5
\NAMES

\RADIO Double rim.
\DATE 18 Feb 2009

\LONGITUDE 16.4 \LATITUDE -0.5
\RAHMS 18 22 38 \DECDM -14 55
\SIZE 13 \TYPE S
\FLUX1GHZ 4.6 \ALPHA 0.7
\NAMES

\NOTES Has been called G16.41$-$0.55.
\RADIO Partial shell.
\DATE 26 Mar 2009

\LONGITUDE 16.7 \LATITUDE +0.1
\RAHMS 18 20 56 \DECDM -14 20
\SIZE 4 \TYPE C
\FLUX1GHZ 3.0 \ALPHA 0.6
\NAMES

\NOTES Has been called G16.73$+$0.08.
\RADIO Asymmetric shell with flat-spectrum core.
\XRAY Non-thermal core.
\DATE 18 Feb 2009

\LONGITUDE 16.8 \LATITUDE -1.1
\RAHMS 18 25 20 \DECDM -14 46
\SIZE 30\X/24? \TYPE ?
\FLUX1GHZ 2? \ALPHA ?
\NAMES

\NOTES Has been called G16.85$-$1.05.
\RADIO Overlapping thermal and non-thermal emission, parameters uncertain.
\POINT Pulsar within boundary of non-thermal emission.
\DATE 13 Aug 1998

\LONGITUDE 17.0 \LATITUDE -0.0
\RAHMS 18 21 57 \DECDM -14 08
\SIZE 5 \TYPE S
\FLUX1GHZ 0.5 \ALPHA 0.5
\NAMES

\NOTES Has been called G17.02$-$0.04.
\RADIO Shell.
\DATE 2006 March 29

\LONGITUDE 17.4 \LATITUDE -2.3
\RAHMS 18 30 55 \DECDM -14 52
\SIZE 24? \TYPE S
\FLUX1GHZ 4.8? \ALPHA 0.8?
\NAMES

\RADIO Incomplete, poorly defined shell.
\OPTICAL Filaments to SE, and diffuse emission.
\DATE 7 Jan 2004

\LONGITUDE 17.4 \LATITUDE -0.1
\RAHMS 18 23 08 \DECDM -13 46
\SIZE 6 \TYPE S
\FLUX1GHZ 0.4 \ALPHA 0.7
\NAMES

\NOTES Has been called G17.48$-$0.12.
\RADIO Partial shell.
\DATE 2006 March 29

\LONGITUDE 17.8 \LATITUDE -2.6
\RAHMS 18 32 50 \DECDM -14 39
\SIZE 24 \TYPE S
\FLUX1GHZ 4.0? \ALPHA 0.3?
\NAMES

\RADIO Well defined shell.
\DATE 13 Aug 1998

\LONGITUDE 18.1 \LATITUDE -0.1
\RAHMS 18 24 34 \DECDM -13 11
\SIZE 8 \TYPE S
\FLUX1GHZ 4.6 \ALPHA 0.5
\NAMES

\NOTES Has been called G18.1$-$0.2 and G18.16$-$0.16.
\RADIO Shell.
\XRAY Detected.

\DATE 2006 March 29

\LONGITUDE 18.6 \LATITUDE -0.2
\RAHMS 18 25 55 \DECDM -12 50
\SIZE 6 \TYPE S
\FLUX1GHZ 1.4 \ALPHA 0.4
\NAMES

\NOTES Has been called G18.62$-$0.28.
\RADIO Partial shell.
\DATE 2006 March 29

\LONGITUDE 18.8 \LATITUDE +0.3
\RAHMS 18 23 58 \DECDM -12 23
\SIZE 17\X/11 \TYPE S
\FLUX1GHZ 33 \ALPHA 0.4
\NAMES Kes 67

\NOTES Has been called G18.9$+$0.3.
\RADIO Incomplete shell, in complex region near the H\II/ region W39.
\DISTANCE Association with molecular cloud and H\I/ absorption suggests 12~kpc.
\DATE 30 Mar 2009

\LONGITUDE 18.9 \LATITUDE -1.1
\RAHMS 18 29 50 \DECDM -12 58
\SIZE 33 \TYPE C?
\FLUX1GHZ 37 \ALPHA varies
\NAMES

\NOTES Has been called G18.95$-$1.1 and G18.94$-$1.04.
\RADIO Non-thermal, diffuse partially limb-brightened, with central ridge.
\XRAY Partial shell.
\DISTANCE Various observations suggest 2~kpc.
\DATE 26 Mar 2009

\LONGITUDE 19.1 \LATITUDE +0.2
\RAHMS 18 24 56 \DECDM -12 07
\SIZE 27 \TYPE S
\FLUX1GHZ 10 \ALPHA 0.5
\NAMES

\NOTES Has been called G19.15$+$0.27.
\RADIO Partial shell.
\DATE 2006 March 29

\LONGITUDE 20.0 \LATITUDE -0.2
\RAHMS 18 28 07 \DECDM -11 35
\SIZE 10 \TYPE F
\FLUX1GHZ 10 \ALPHA 0.0
\NAMES

\RADIO Faint, filled-centre, polarised.
\POINT OH source 20.1$-$0.1 is nearby.
\DATE 13 Aug 1998

\LONGITUDE 20.4 \LATITUDE +0.1
\RAHMS 18 27 51 \DECDM -11 00
\SIZE 8 \TYPE S
\FLUX1GHZ 3.1 \ALPHA 0.4
\NAMES

\NOTES Has been called G20.47$+$0.16.
\RADIO Shell.
\DATE 2006 March 29

\LONGITUDE 21.0 \LATITUDE -0.4
\RAHMS 18 31 12 \DECDM -10 47
\SIZE 9\X/7 \TYPE S
\FLUX1GHZ 1.1 \ALPHA 0.6
\NAMES

\NOTES Has been called G21.04$-$0.47.
\RADIO Shell.
\DATE 2006 March 29

\LONGITUDE 21.5 \LATITUDE -0.9
\RAHMS 18 33 33 \DECDM -10 35
\SIZE 4 \TYPE C
\FLUX1GHZ 6? \ALPHA 0.0
\NAMES

\NOTES Early observations relate to the central core only.
\RADIO Filled-centre, with high frequency turnover.
\XRAY Central core, with extended, faint halo.
\POINT Central pulsar.
\DISTANCE H\I/ absorption indicates 4.6~kpc.
\DATE 18 Feb 2009

\LONGITUDE 21.5 \LATITUDE -0.1
\RAHMS 18 30 50 \DECDM -10 09
\SIZE 5 \TYPE S
\FLUX1GHZ 0.4 \ALPHA 0.5
\NAMES

\NOTES Has been called G21.56$-$0.10.
\RADIO Partial shell.
\XRAY Detected.
\DATE 2006 March 29

\LONGITUDE 21.8 \LATITUDE -0.6
\RAHMS 18 32 45 \DECDM -10 08
\SIZE 20 \TYPE S
\FLUX1GHZ 69 \ALPHA 0.5
\NAMES Kes 69

\RADIO Incomplete shell.
\XRAY Detected.
\DISTANCE H\I/ absorption indicates 5.5 to 7.4~kpc.
\DATE 16 Mar 2009

\LONGITUDE 22.7 \LATITUDE -0.2
\RAHMS 18 33 15 \DECDM -09 13
\SIZE 26 \TYPE S?
\FLUX1GHZ 33 \ALPHA 0.6
\NAMES

\RADIO Non-thermal ring in complex region, overlapping G23.3$-$0.3.
\DATE 16 Mar 2009

\LONGITUDE 23.3 \LATITUDE -0.3
\RAHMS 18 34 45 \DECDM -08 48
\SIZE 27 \TYPE S
\FLUX1GHZ 70 \ALPHA 0.5
\NAMES W41

\RADIO Distorted ring, in complex region, overlapping G22.7$-$0.2.
\XRAY Possible extended emission, with compact sources.
\POINT Pulsar association suggested.
\DISTANCE H\I/ and CO observations indicate 4.2~kpc.
\DATE 16 Mar 2009

\LONGITUDE 23.6 \LATITUDE +0.3
\RAHMS 18 33 03 \DECDM -08 13
\SIZE 10? \TYPE ?
\FLUX1GHZ 8? \ALPHA 0.3
\NAMES

\RADIO Not well resolved, in complex region.
\DATE 13 Aug 1998

\LONGITUDE 24.7 \LATITUDE -0.6
\RAHMS 18 38 43 \DECDM -07 32
\SIZE 15? \TYPE S?
\FLUX1GHZ 8 \ALPHA 0.5
\NAMES

\RADIO Incomplete shell, defined in SW.
\DATE 13 Aug 1998

\LONGITUDE 24.7 \LATITUDE +0.6
\RAHMS 18 34 10 \DECDM -07 05
\SIZE 30\X/15 \TYPE C?
\FLUX1GHZ 20? \ALPHA 0.2?
\NAMES

\RADIO Filled-centre, with faint shell, and a compact H\II/ region to the S.
\DATE 13 Aug 1998

\LONGITUDE 27.4 \LATITUDE +0.0
\RAHMS 18 41 19 \DECDM -04 56
\SIZE 4 \TYPE S
\FLUX1GHZ 6 \ALPHA 0.68
\NAMES 4C$-$04.71

\NOTES Early references refer to G27.3$-$0.1 (Kes 73), a supposed larger remnant.
\RADIO Incomplete shell.
\XRAY Diffuse emission, with central low period pulsar.
\POINT Central AXP.
\DISTANCE H\I/ absorption suggests 7.5 to 9.8~kpc.
\DATE 18 Feb 2009

\LONGITUDE 27.8 \LATITUDE +0.6
\RAHMS 18 39 50 \DECDM -04 24
\SIZE 50\X/30 \TYPE F
\FLUX1GHZ 30 \ALPHA varies
\NAMES

\RADIO Filled-centre, with spectral turnover.
\DATE 18 Feb 2009

\LONGITUDE 28.6 \LATITUDE -0.1
\RAHMS 18 43 55 \DECDM -03 53
\SIZE 13\X/9 \TYPE S
\FLUX1GHZ 3? \ALPHA ?
\NAMES

\RADIO Poorly defined regions of non-thermal emission.
\XRAY Diffuse shell, with thermal and non-thermal emission.
\DATE 11 Jan 2004

\LONGITUDE 28.8 \LATITUDE +1.5
\RAHMS 18 39 00 \DECDM -02 55
\SIZE 100? \TYPE S?
\FLUX1GHZ ? \ALPHA 0.4?
\NAMES

\RADIO Part of rim detected.
\XRAY Diffuse, Centrally brightened.
\DATE 18 Feb 2009

\LONGITUDE 29.6 \LATITUDE +0.1
\RAHMS 18 44 52 \DECDM -02 57
\SIZE 5 \TYPE S
\FLUX1GHZ 1.5? \ALPHA 0.5?
\NAMES

\RADIO Diffuse shell.
\POINT AXP associated.
\DATE 10 Oct 2001

\LONGITUDE 29.7 \LATITUDE -0.3
\RAHMS 18 46 25 \DECDM -02 59
\SIZE 3 \TYPE C
\FLUX1GHZ 10 \ALPHA 0.7
\NAMES Kes 75

\NOTES Has erroneously been called G29.6$+$0.1.
\RADIO Shell with flatter spectrum emission from centre.
\XRAY Thermal shell and non-thermal core, and central pulsar.
\POINT X-ray pulsar.
\DISTANCE H\I/ absorption indicates possibly $<7.5$~kpc.
\DATE 19 Feb 2009

\LONGITUDE 30.7 \LATITUDE -2.0
\RAHMS 18 54 25 \DECDM -02 54
\SIZE 16 \TYPE ?
\FLUX1GHZ 0.5? \ALPHA 0.7?
\NAMES

\RADIO Poorly defined.
\DATE 13 Aug 1998

\LONGITUDE 30.7 \LATITUDE +1.0
\RAHMS 18 44 00 \DECDM -01 32
\SIZE 24\X/18 \TYPE S?
\FLUX1GHZ 6 \ALPHA 0.4
\NAMES

\RADIO Non-thermal, highly polarised part shell?
\POINT Compact source near centre.
\DATE 13 Aug 1998

\LONGITUDE 31.5 \LATITUDE -0.6
\RAHMS 18 51 10 \DECDM -01 31
\SIZE 18? \TYPE S?
\FLUX1GHZ 2? \ALPHA ?
\NAMES

\NOTES Has been called G31.55$-$0.65.
\RADIO Distorted shell? near H\II/ region.
\OPTICAL Diffuse, incomplete shell.
\DATE 17 Oct 2001

\LONGITUDE 31.9 \LATITUDE +0.0
\RAHMS 18 49 25 \DECDM -00 55
\SIZE 7\X/5 \TYPE S
\FLUX1GHZ 24 \ALPHA 0.49
\NAMES 3C391

\RADIO Shell, brightest in NW, with low frequency turnover.
\XRAY Diffuse with central core.
\DISTANCE H\I/ absorption is seen to the tangent point (8.5~kpc).
\DATE 16 Mar 2009

\LONGITUDE 32.0 \LATITUDE -4.9
\RAHMS 19 06 00 \DECDM -03 00
\SIZE 60? \TYPE S?
\FLUX1GHZ 22? \ALPHA 0.5?
\NAMES 3C396.1

\RADIO Possible large shell?
\DATE 25th February 1988

\LONGITUDE 32.1 \LATITUDE -0.9
\RAHMS 18 53 10 \DECDM -01 08
\SIZE 40? \TYPE C?
\FLUX1GHZ ? \ALPHA ?
\NAMES

\RADIO Possible faint shell, not well defined.
\XRAY Diffuse, with clumps.
\DATE 17 Aug 1998

\LONGITUDE 32.4 \LATITUDE +0.1
\RAHMS 18 50 05 \DECDM -00 25
\SIZE 6 \TYPE S
\FLUX1GHZ 0.25? \ALPHA ?
\NAMES

\NOTES Has been called G32.45$+$0.1.
\RADIO Shell.
\XRAY Shell.
\DISTANCE X-ray absorption suggests 17~kpc.
\DATE 16 Mar 2009

\LONGITUDE 32.8 \LATITUDE -0.1
\RAHMS 18 51 25 \DECDM -00 08
\SIZE 17 \TYPE S?
\FLUX1GHZ 11? \ALPHA 0.2?
\NAMES Kes 78

\NOTES Part has been called G33.1$-$0.1.
\RADIO Elongated shell?
\DATE 1 Aug 2000

\LONGITUDE 33.2 \LATITUDE -0.6
\RAHMS 18 53 50 \DECDM -00 02
\SIZE 18 \TYPE S
\FLUX1GHZ 3.5 \ALPHA varies
\NAMES

\RADIO Incomplete shell.
\DATE 13 Aug 1998

\LONGITUDE 33.6 \LATITUDE +0.1
\RAHMS 18 52 48 \DECDM +00 41
\SIZE 10 \TYPE S
\FLUX1GHZ 22 \ALPHA 0.5
\NAMES Kes 79, 4C00.70, HC13

\NOTES Has been called G33.7$+$0.0.
\RADIO Shell, with bright central region, in complex region.
\XRAY Multiple shells and filaments.
\POINT Central X-ray pulsar.
\DISTANCE H\I/ absorption gives about 7.8~kpc.
\DATE 16 Mar 2009

\LONGITUDE 34.7 \LATITUDE -0.4
\RAHMS 18 56 00 \DECDM +01 22
\SIZE 35\X/27 \TYPE C
\FLUX1GHZ 230 \ALPHA 0.37
\NAMES W44, 3C392

\NOTES Has been called G34.6$-$0.5.
\RADIO Distorted shell, brighter to the E, with pulsar and associated nebula.
\OPTICAL Diffuse emission.
\XRAY Centrally concentrated, thermal spectrum, plus pulsar wind nebula.
\POINT Pulsar within the boundary of the remnant.
\DISTANCE H\I/ absorption indicates 2.8~kpc.
\DATE 16 Mar 2009

\LONGITUDE 36.6 \LATITUDE -0.7
\RAHMS 19 00 35 \DECDM +02 56
\SIZE 25? \TYPE S?
\FLUX1GHZ ? \ALPHA ?
\NAMES

\RADIO polarised arc, possibly part of a larger shell?
\DATE 16 Mar 2009

\LONGITUDE 36.6 \LATITUDE +2.6
\RAHMS 18 48 49 \DECDM +04 26
\SIZE 17\X/13? \TYPE S
\FLUX1GHZ 0.7? \ALPHA 0.5?
\NAMES

\RADIO Poorly resolved shell.
\DATE 19th May 1992

\LONGITUDE 39.2 \LATITUDE -0.3
\RAHMS 19 04 08 \DECDM +05 28
\SIZE 8\X/6 \TYPE C
\FLUX1GHZ 18 \ALPHA 0.6
\NAMES 3C396, HC24, NRAO 593

\RADIO Shell, brighter to W, with faint `tail' to E.
\XRAY Diffuse, brighter to W, with central core.
\POINT Central X-ray source.
\DISTANCE H\I/ absorption suggests $>7.7$~kpc.
\DATE 16 Mar 2009

\LONGITUDE 39.7 \LATITUDE -2.0
\RAHMS 19 12 20 \DECDM +04 55
\SIZE 120\X/60 \TYPE ?
\FLUX1GHZ 85? \ALPHA 0.7?
\NAMES W50, SS433

\NOTES Eastern part has been called G40.0$-$3.1. Is this a SNR?
\RADIO Elongated shell, containing SS433, adjacent to the H\II/ region S74.
\OPTICAL Faint filaments at the edge of the radio emission.
\XRAY Emission from SS433 and two lobes.
\POINT SS433 is the compact source in the centre of the W50.
\DISTANCE H\I/ absorption indicates $6.0\PM/0.5$~kpc.
\DATE 19 Feb 2009

\LONGITUDE 40.5 \LATITUDE -0.5
\RAHMS 19 07 10 \DECDM +06 31
\SIZE 22 \TYPE S
\FLUX1GHZ 11 \ALPHA 0.5
\NAMES

\RADIO Shell, brightest to the NE.
\DATE 19 Feb 2009

\LONGITUDE 41.1 \LATITUDE -0.3
\RAHMS 19 07 34 \DECDM +07 08
\SIZE 4.5\X/2.5 \TYPE S
\FLUX1GHZ 22 \ALPHA 0.48
\NAMES 3C397

\RADIO 3C397 is two sources: the E is the SNR, the W is a H\II/ region.
\XRAY Brighter to the E and W, with central component.
\DISTANCE Possible limit of $>7.5$~kpc for non-thermal component from H\I/ absorption.
\DATE 16 Mar 2009

\LONGITUDE 42.8 \LATITUDE +0.6
\RAHMS 19 07 20 \DECDM +09 05
\SIZE 24 \TYPE S
\FLUX1GHZ 3? \ALPHA 0.5?
\NAMES

\NOTES Has been called G42.8$+$0.65.
\RADIO Faint shell.
\POINT Near soft gamma repeater, and young pulsar.
\DATE 11 Jan 2004

\LONGITUDE 43.3 \LATITUDE -0.2
\RAHMS 19 11 08 \DECDM +09 06
\SIZE 4\X/3 \TYPE S
\FLUX1GHZ 38 \ALPHA 0.48
\NAMES W49B

\RADIO Shell, brightest to the SE and W, near the H\II/ region W49A.
\XRAY Centrally brightened, elongated E--W.
\DISTANCE H\I/ absorption indicates 10~kpc.
\DATE 16 Mar 2009

\LONGITUDE 43.9 \LATITUDE +1.6
\RAHMS 19 05 50 \DECDM +10 30
\SIZE 60? \TYPE S?
\FLUX1GHZ 8.6? \ALPHA 0.2?
\NAMES

\RADIO Large, poorly defined faint shell.
\POINT Soft gamma repeater nearby.
\DATE 11 Jan 2004

\LONGITUDE 45.7 \LATITUDE -0.4
\RAHMS 19 16 25 \DECDM +11 09
\SIZE 22 \TYPE S
\FLUX1GHZ 4.2? \ALPHA 0.4?
\NAMES

\RADIO Shell, brightest to the SE, poorly defined to NW.
\DATE 16 Mar 2009

\LONGITUDE 46.8 \LATITUDE -0.3
\RAHMS 19 18 10 \DECDM +12 09
\SIZE 17\X/13 \TYPE S
\FLUX1GHZ 14 \ALPHA 0.5
\NAMES (HC30)

\NOTES Has been called G46.6$-$0.2.
\RADIO Shell, two bright arcs to NNW and SSE.
\DISTANCE H\I/ absorption suggests 6.8--8.8~kpc.
\DATE 31 Mar 2006

\LONGITUDE 49.2 \LATITUDE -0.7
\RAHMS 19 23 50 \DECDM +14 06
\SIZE 30 \TYPE S?
\FLUX1GHZ 160? \ALPHA 0.3?
\NAMES (W51)

\RADIO In complex region, parameters uncertain.
\XRAY Elongated east--west.
\OPTICAL Some diffuse emission possibly associated.
\DISTANCE Association with CO gives 6~kpc.
\DATE 19 Feb 2009

\LONGITUDE 53.6 \LATITUDE -2.2
\RAHMS 19 38 50 \DECDM +17 14
\SIZE 33\X/28 \TYPE S
\FLUX1GHZ 8 \ALPHA 0.75
\NAMES 3C400.2, NRAO 611

\NOTES Has been called G53.7$-$2.2.
\RADIO Ring of emission, with extension to NW.
\OPTICAL Filaments and diffuse emission.
\XRAY Centrally brightened, offset to NW.
\DISTANCE Association with H\I/ gives 2.8~kpc.
\DATE 19 Feb 2009

\LONGITUDE 54.1 \LATITUDE +0.3
\RAHMS 19 30 31 \DECDM +18 52
\SIZE 1.5 \TYPE F?
\FLUX1GHZ 0.5 \ALPHA 0.1
\NAMES

\RADIO Filled-centre.
\XRAY Centrally concentrated, with extensions and diffuse emission.
\POINT Central pulsar.
\DISTANCE HI absorption suggests 4.5--9~kpc, association with CO suggest 8.2~kpc.
\DATE 19 Feb 2009

\LONGITUDE 54.4 \LATITUDE -0.3
\RAHMS 19 33 20 \DECDM +18 56
\SIZE 40 \TYPE S
\FLUX1GHZ 28 \ALPHA 0.5
\NAMES (HC40)

\NOTES Has been called G54.5$-$0.3.
\RADIO Shell, in complex region.
\OPTICAL Faint filaments.
\DATE 16 Mar 2009

\LONGITUDE 55.0 \LATITUDE +0.3
\RAHMS 19 32 00 \DECDM +19 50
\SIZE 20\X/15? \TYPE S
\FLUX1GHZ 0.5? \ALPHA 0.5?
\NAMES

\NOTES Has been called G55.2$+$0.5.
\RADIO Faint, partial shell.
\DISTANCE Association with H\I/ features implies 14~kpc.
\POINT Old pulsar nearby.
\DATE 16 Mar 2009

\LONGITUDE 55.7 \LATITUDE +3.4
\RAHMS 19 21 20 \DECDM +21 44
\SIZE 23 \TYPE S
\FLUX1GHZ 1.4 \ALPHA 0.6
\NAMES

\RADIO Incomplete shell.
\POINT Old pulsar within the boundary of the remnant.
\DATE 05 Apr 1993

\LONGITUDE 57.2 \LATITUDE +0.8
\RAHMS 19 34 59 \DECDM +21 57
\SIZE 12? \TYPE S?
\FLUX1GHZ 1.8? \ALPHA ?
\NAMES (4C21.53)

\RADIO Extended non-thermal arc.
\POINT Near the millisecond pulsar, but not thought to be related.
\DATE 1st March 1988

\LONGITUDE 59.5 \LATITUDE +0.1
\RAHMS 19 42 33 \DECDM +23 35
\SIZE 15 \TYPE S
\FLUX1GHZ 3? \ALPHA ?
\NAMES

\NOTES Has been called G59.6$+$0.1.
\RADIO Incomplete shell.
\OPTICAL Diffuse shell.
\DATE 19 Feb 2009

\LONGITUDE 59.8 \LATITUDE +1.2
\RAHMS 19 38 55 \DECDM +24 19
\SIZE 20\X/16? \TYPE ?
\FLUX1GHZ 1.6 \ALPHA 0.5
\NAMES

\NOTES Has been called G59.7$+$1.2.
\RADIO Poorly defined source.
\OPTICAL Faint diffuse emission and filaments.
\DATE 31 Mar 2006

\LONGITUDE 63.7 \LATITUDE +1.1
\RAHMS 19 47 52 \DECDM +27 45
\SIZE 8 \TYPE F
\FLUX1GHZ 1.8 \ALPHA 0.3
\NAMES

\RADIO Centrally brightened, with core.
\DATE 13 Aug 1998

\LONGITUDE 65.1 \LATITUDE +0.6
\RAHMS 19 54 40 \DECDM +28 35
\SIZE 90\X/50 \TYPE S
\FLUX1GHZ 5.5 \ALPHA 0.61
\NAMES

\RADIO Large, faint shell.
\POINT Old pulsar nearby.
\DISTANCE Possible association with H\I/ suggests 9~kpc.
\DATE 20 Feb 2009

\LONGITUDE 65.3 \LATITUDE +5.7
\RAHMS 19 33 00 \DECDM +31 10
\SIZE 310\X/240 \TYPE S?
\FLUX1GHZ 52? \ALPHA 0.6?
\NAMES

\NOTES Has been called G65.2$+$5.7.
\RADIO Large, faint ring? near S91 and S94.
\OPTICAL Filamentary ring.
\XRAY Diffuse, centrally brightened.
\DISTANCE Optical proper motions and velocities indicates 0.8~kpc.
\DATE 19 Feb 2009

\LONGITUDE 65.7 \LATITUDE +1.2
\RAHMS 19 52 10 \DECDM +29 26
\SIZE 22 \TYPE F
\FLUX1GHZ 5.1 \ALPHA varies
\NAMES DA 495

\NOTES Has mistakenly been called G55.7$+$1.2.
\RADIO Centrally brightened with thick shell?
\XRAY Detected.
\POINT Compact X-ray source near centre.
\DISTANCE H\I/ polarisation observations suggest 1.5~kpc.
\DATE 30 Mar 2009

\LONGITUDE 67.7 \LATITUDE +1.8
\RAHMS 19 54 32 \DECDM +31 29
\SIZE 15\X/12 \TYPE S
\FLUX1GHZ 1.0 \ALPHA 0.5
\NAMES

\RADIO Double arc shell.
\OPTICAL Filaments in N.
\DATE 20 Feb 2009

\LONGITUDE 68.6 \LATITUDE -1.2
\RAHMS 20 08 40 \DECDM +30 37
\SIZE 23 \TYPE ?
\FLUX1GHZ 0.7? \ALPHA 0.0?
\NAMES

\RADIO Faint, poorly defined source.
\DATE 20 Feb 2009

\LONGITUDE 69.0 \LATITUDE +2.7
\RAHMS 19 53 20 \DECDM +32 55
\SIZE 80? \TYPE ?
\FLUX1GHZ 120? \ALPHA varies
\NAMES CTB 80

\NOTES An association with a SN in \AD/1408 has been suggested. Has been called G68.8$+$2.8. Is it a SNR?
\RADIO Compact core, flat spectrum plateau, and steeper spectrum extensions, with spectral break?
\OPTICAL Expanding nebulosity near centre, with filaments to the SW and far NE.
\XRAY Diffuse emission with compact source.
\POINT Pulsar at western edge of core.
\DATE 20 Feb 2009

\LONGITUDE 69.7 \LATITUDE +1.0
\RAHMS 20 02 40 \DECDM +32 43
\SIZE 16\X/14 \TYPE S
\FLUX1GHZ 2.0 \ALPHA 0.7
\NAMES

\RADIO Poorly resolved source.
\XRAY Detected.
\DATE 20 Feb 2009

\LONGITUDE 73.9 \LATITUDE +0.9
\RAHMS 20 14 15 \DECDM +36 12
\SIZE 27 \TYPE S?
\FLUX1GHZ 9 \ALPHA 0.23
\NAMES

\RADIO Diffuse, centrally brightened to SW.
\OPTICAL Faint shell.
\DATE 20 Feb 2009

\LONGITUDE 74.0 \LATITUDE -8.5
\RAHMS 20 51 00 \DECDM +30 40
\SIZE 230\X/160 \TYPE S
\FLUX1GHZ 210 \ALPHA varies
\NAMES Cygnus Loop

\NOTES Has been suggested that this is two overlapping remnants.
\RADIO Shell, brightest to the NE, with fainter breakout region to S, with spectral variations.
\OPTICAL Large filamentary loop, brightest to the NE, not well defined to the S or W.
\XRAY Shell in soft X-rays.
\POINT Several compact radio sources within the boundary of the remnant, including CL4, plus X-ray sources in S.
\DISTANCE Optical proper motion and shock velocity gives 0.44~kpc.
\DATE 19 Feb 2009

\LONGITUDE 74.9 \LATITUDE +1.2
\RAHMS 20 16 02 \DECDM +37 12
\SIZE 8\X/6 \TYPE F
\FLUX1GHZ 9 \ALPHA varies
\NAMES CTB 87

\RADIO Filled-centre, with high polarisation and high frequency turnover.
\XRAY Centrally brightened.
\DISTANCE H\I/ absorption indicates 12~kpc, optical extinction gives 6.1~kpc.
\POINT Extragalactic compact source is nearby.
\DATE 20 Feb 2009

\LONGITUDE 76.9 \LATITUDE +1.0
\RAHMS 20 22 20 \DECDM +38 43
\SIZE 9 \TYPE ?
\FLUX1GHZ 1.2 \ALPHA 0.60
\NAMES

\RADIO Diffuse, non-thermal, with low frequency turnover.
\DATE 20 Feb 2009

\LONGITUDE 78.2 \LATITUDE +2.1
\RAHMS 20 20 50 \DECDM +40 26
\SIZE 60 \TYPE S
\FLUX1GHZ 320 \ALPHA 0.51
\NAMES DR4, $\gamma$ Cygni SNR

\NOTES Has been called G78.1$+$1.8.
\RADIO In complex region (early catalogues refer to other proposed remnants in this region).
\OPTICAL Faint filaments, spectra indicate a SNR superposed on a H\II/ region.
\XRAY Weak emission from the SE of the remnant.
\POINT $\gamma$-ray and X-ray point source in remnant.
\DATE 20 Feb 2009

\LONGITUDE 82.2 \LATITUDE +5.3
\RAHMS 20 19 00 \DECDM +45 30
\SIZE 95\X/65 \TYPE S
\FLUX1GHZ 120? \ALPHA 0.5?
\NAMES W63

\NOTES Has been called G82.5$+$5.3.
\RADIO Shell in the Cygnus X complex.
\OPTICAL In complex region, but spectra indicate SNR filaments.
\XRAY Detected.
\DATE 20 Feb 2009

\LONGITUDE 83.0 \LATITUDE -0.3
\RAHMS 20 46 55 \DECDM +42 52
\SIZE 9\X/7 \TYPE S
\FLUX1GHZ 1 \ALPHA 0.4
\NAMES

\RADIO Incomplete shell.
\DATE 16 Mar 2009

\LONGITUDE 84.2 \LATITUDE -0.8
\RAHMS 20 53 20 \DECDM +43 27
\SIZE 20\X/16 \TYPE S
\FLUX1GHZ 11 \ALPHA 0.5
\NAMES

\RADIO Elongated shell, with a filament aligned with the major axis.
\DATE 20 Feb 2009

\LONGITUDE 85.4 \LATITUDE +0.7
\RAHMS 20 50 40 \DECDM +45 22
\SIZE 24? \TYPE S
\FLUX1GHZ ? \ALPHA 0.2
\NAMES

\RADIO Faint, incomplete shell, within larger thermal shell.
\XRAY Centrally brightened.
\DISTANCE H\I/ observations suggest 3.5~kpc.
\DATE 20 Feb 2009

\LONGITUDE 85.9 \LATITUDE -0.6
\RAHMS 20 58 40 \DECDM +44 53
\SIZE 24 \TYPE S
\FLUX1GHZ ? \ALPHA 0.2
\NAMES

\RADIO Faint, incomplete shell.
\XRAY Centrally brightened.
\DISTANCE H\I/ observations suggest 4.8~kpc.
\DATE 20 Feb 2009

\LONGITUDE 89.0 \LATITUDE +4.7
\RAHMS 20 45 00 \DECDM +50 35
\SIZE 120\X/90 \TYPE S
\FLUX1GHZ 220 \ALPHA 0.38
\NAMES HB21

\RADIO Distorted shell (4C50.52, an extragalactic double, is within the boundary of the remnant).
\OPTICAL Filaments and patches.
\XRAY Centrally brightened.
\DISTANCE Various associations imply 0.8~kpc.
\DATE 16 Mar 2009

\LONGITUDE 93.3 \LATITUDE +6.9
\RAHMS 20 52 25 \DECDM +55 21
\SIZE 27\X/20 \TYPE C?
\FLUX1GHZ 9 \ALPHA 0.45
\NAMES DA 530, 4C(T)55.38.1

\NOTES Has been called G93.2$+$6.7.
\RADIO Shell, with two bright limbs, highly polarised.
\XRAY Compact central source.
\DISTANCE H\I/ observations suggest 2.2~kpc.
\DATE 19 Feb 2009

\LONGITUDE 93.7 \LATITUDE -0.2
\RAHMS 21 29 20 \DECDM +50 50
\SIZE 80 \TYPE S
\FLUX1GHZ 65 \ALPHA 0.65
\NAMES CTB 104A, DA 551

\NOTES Has been called G93.6$-$0.2 and G93.7$-$0.3.
\RADIO Distorted, faint shell.
\DISTANCE Association with H\I/ features suggests 1.5~kpc.
\DATE 20 Feb 2009

\LONGITUDE 94.0 \LATITUDE +1.0
\RAHMS 21 24 50 \DECDM +51 53
\SIZE 30\X/25 \TYPE S
\FLUX1GHZ 13 \ALPHA 0.48
\NAMES 3C434.1

\RADIO Incomplete shell, containing H\I/ shell.
\DISTANCE Association with stellar wind bubble implies 5.2~kpc.
\DATE 20 Feb 2009

\LONGITUDE 96.0 \LATITUDE +2.0
\RAHMS 21 30 30 \DECDM +53 59
\SIZE 26 \TYPE S
\FLUX1GHZ 0.3 \ALPHA 0.5
\NAMES

\RADIO Faint, arc in S, poorly defined in N.
\DISTANCE Association for H\I/ indicates 4~kpc.
\DATE 2006 March 29

\LONGITUDE 106.3 \LATITUDE +2.7
\RAHMS 22 27 30 \DECDM +60 50
\SIZE 60\X/24 \TYPE C?
\FLUX1GHZ 6 \ALPHA 0.6
\NAMES

\NOTES Incorporates the pulsar wind nebula G106.6$+$2.9 (the `Boomerang').
\RADIO Faint extended source, which brighter `head' to NE.
\XRAY Pulsar and wind nebula.
\POINT Pulsar.
\DATE 20 Feb 2009

\LONGITUDE 108.2 \LATITUDE -0.6
\RAHMS 22 53 40 \DECDM +58 50
\SIZE 70\X/54 \TYPE S
\FLUX1GHZ 8 \ALPHA 0.5
\NAMES

\RADIO Faint shell.
\DISTANCE Possible associated H\I/ structures suggest 3.2~kpc.
\DATE 16 Mar 2009

\LONGITUDE 109.1 \LATITUDE -1.0
\RAHMS 23 01 35 \DECDM +58 53
\SIZE 28 \TYPE S
\FLUX1GHZ 22 \ALPHA 0.50
\NAMES CTB 109

\RADIO Semicircular shell, with the Molecular cloud S152 is to the immediate W.
\XRAY Semicircular shell, with pulsar at W edge.
\POINT Long period X-ray pulsar.
\DISTANCE Association with H\II/ regions implies 3~kpc.
\DATE 20 Feb 2009

\LONGITUDE 111.7 \LATITUDE -2.1
\RAHMS 23 23 26 \DECDM +58 48
\SIZE 5 \TYPE S
\FLUX1GHZ 2720 \ALPHA 0.77
\NAMES Cassiopeia A, 3C461

\NOTES Presumably the remnant of a late 17th century SN.
\RADIO Bright shell with compact knots and extended plateau of emission.
\OPTICAL Fast knots and quasi-stationary flocculli, with many filaments at large radii, and NE `jet'.
\XRAY Incomplete shell, with hard spectral component, and compact central source.
\DISTANCE Optical expansion, plus proper motions indicate 3.4~kpc.
\DATE 19 Feb 2009

\LONGITUDE 113.0 \LATITUDE +0.2
\RAHMS 23 36 35 \DECDM +61 22
\SIZE 40\X/17? \TYPE ?
\FLUX1GHZ ? \ALPHA ?
\NAMES

\RADIO Elongated, extent not well defined.
\DISTANCE Association for H\I/ indicates 3.1~kpc.
\POINT Contains old pulsar.
\DATE 16 Mar 2009

\LONGITUDE 114.3 \LATITUDE +0.3
\RAHMS 23 37 00 \DECDM +61 55
\SIZE 90\X/55 \TYPE S
\FLUX1GHZ 5.5 \ALPHA 0.5
\NAMES

\RADIO Shell, with H\II/ region S165 within the boundary of the remnant.
\OPTICAL Faint emission in centre and to S.
\DISTANCE Association with H\I/ and other features implies 0.7~kpc.
\POINT Pulsar near centre of remnant.
\DATE 20 Feb 2009

\LONGITUDE 116.5 \LATITUDE +1.1
\RAHMS 23 53 40 \DECDM +63 15
\SIZE 80\X/60 \TYPE S
\FLUX1GHZ 10 \ALPHA 0.5
\NAMES

\RADIO Distinct shell, with high polarisation.
\OPTICAL Detected.
\DISTANCE Association with H\I/ features implies 1.6~kpc.
\DATE 20 Feb 2009

\LONGITUDE 116.9 \LATITUDE +0.2
\RAHMS 23 59 10 \DECDM +62 26
\SIZE 34 \TYPE S
\FLUX1GHZ 8 \ALPHA 0.61
\NAMES CTB 1

\NOTES Has been called G117.3$+$0.1 and G116.9$+$0.1.
\RADIO Incomplete shell.
\OPTICAL Filaments on sky survey.
\XRAY Centrally brightened, with NE `breakout'.
\POINT Pulsar to NE.
\DISTANCE Association with H\I/ features implies 1.6~kpc.
\DATE 20 Feb 2009

\LONGITUDE 119.5 \LATITUDE +10.2
\RAHMS 00 06 40 \DECDM +72 45
\SIZE 90? \TYPE S
\FLUX1GHZ 36 \ALPHA 0.6
\NAMES CTA 1

\NOTES Has been called G119.5$+$10.3.
\RADIO Incomplete shell, with `breakout' to NW.
\OPTICAL Faint diffuse nebulosities.
\XRAY Centrally brightened.
\POINT $\gamma$-ray pulsar.
\DISTANCE Associated H\I/ shell indicates 1.4~kpc.
\DATE 19 Feb 2009

\LONGITUDE 120.1 \LATITUDE +1.4
\RAHMS 00 25 18 \DECDM +64 09
\SIZE 8 \TYPE S
\FLUX1GHZ 56 \ALPHA 0.65
\NAMES Tycho, 3C10, SN1572

\NOTES This is the remnant of the Tycho's SN of \AD/1572.
\RADIO Shell, brightest to the NE.
\OPTICAL Faint filaments/knots to the NNW, NE and E.
\XRAY Shell, brighter to the NE.
\POINT Faint radio source near centre of the remnant, thought to be extragalactic.
\DISTANCE H\I/ absorption gives 2--5~kpc, optical proper motion and shock velocity gives 2.4~kpc.
\DATE 20 Feb 2009

\LONGITUDE 126.2 \LATITUDE +1.6
\RAHMS 01 22 00 \DECDM +64 15
\SIZE 70 \TYPE S?
\FLUX1GHZ 6 \ALPHA 0.5
\NAMES

\RADIO Poorly defined shell.
\OPTICAL Filaments, mostly in W.
\DATE 20 Feb 2009

\LONGITUDE 127.1 \LATITUDE +0.5
\RAHMS 01 28 20 \DECDM +63 10
\SIZE 45 \TYPE S
\FLUX1GHZ 12 \ALPHA 0.45
\NAMES R5

\NOTES Has been called G127.3$+$0.7.
\RADIO Distinct shell, with bright central source.
\POINT Flat radio spectrum (extragalactic) source at centre of remnant.
\OPTICAL Detected.
\DISTANCE 1.2--1.3~kpc if associated with NGC~559.
\DATE 20 Feb 2009

\LONGITUDE 130.7 \LATITUDE +3.1
\RAHMS 02 05 41 \DECDM +64 49
\SIZE 9\X/5 \TYPE F
\FLUX1GHZ 33 \ALPHA 0.07
\NAMES 3C58, SN1181

\NOTES This is the remnant of the SN of \AD/1181.
\RADIO Filled-centre, highly polarised, with high frequency turnover.
\OPTICAL Faint filaments.
\XRAY Centrally brightened, with faint jet.
\POINT Central pulsar.
\DISTANCE H\I/ absorption indicates 3.2~kpc.
\DATE 20 Feb 2009

\LONGITUDE 132.7 \LATITUDE +1.3
\RAHMS 02 17 40 \DECDM +62 45
\SIZE 80 \TYPE S
\FLUX1GHZ 45 \ALPHA 0.6
\NAMES HB3

\NOTES Has been called G132.4$+$2.2.
\RADIO Faint shell, adjacent to W3/4/5 complex.
\OPTICAL Complete, filamentary shell, shock excited spectra.
\XRAY Partial shell.
\POINT Pulsar nearby.
\DISTANCE Interaction with surroundings suggests 2.2~kpc.
\DATE 20 Feb 2009

\LONGITUDE 156.2 \LATITUDE +5.7
\RAHMS 04 58 40 \DECDM +51 50
\SIZE 110 \TYPE S
\FLUX1GHZ 5 \ALPHA 0.5
\NAMES

\RADIO Faint shell, brighter in E and W.
\OPTICAL Filamentary ring and smaller patchy ring.
\XRAY Faint shell.
\DATE 20 Feb 2009

\LONGITUDE 160.9 \LATITUDE +2.6
\RAHMS 05 01 00 \DECDM +46 40
\SIZE 140\X/120 \TYPE S
\FLUX1GHZ 110 \ALPHA 0.64
\NAMES HB9

\NOTES Has been called G160.5$+$2.8 and G160.4$+$2.8.
\RADIO Large, filamentary shell.
\OPTICAL Incomplete shell.
\XRAY Centrally brightened.
\POINT Pulsar within boundary of the remnant, plus several nearby compact radio sources.
\DISTANCE Various observations suggests less than 4~kpc.
\DATE 20 Feb 2009

\LONGITUDE 166.0 \LATITUDE +4.3
\RAHMS 05 26 30 \DECDM +42 56
\SIZE 55\X/35 \TYPE S
\FLUX1GHZ 7 \ALPHA 0.37
\NAMES VRO 42.05.01

\RADIO Two arcs of strikingly different radii.
\OPTICAL Nearly complete ring.
\XRAY Predominantly in SW.
\DISTANCE H\I/ indicates 4.5~kpc.
\DATE 19 Feb 2009

\LONGITUDE 179.0 \LATITUDE +2.6
\RAHMS 05 53 40 \DECDM +31 05
\SIZE 70 \TYPE S?
\FLUX1GHZ 7 \ALPHA 0.4
\NAMES

\RADIO Thick shell, with background extragalactic sources near centre.
\DATE 2 Apr 2006

\LONGITUDE 180.0 \LATITUDE -1.7
\RAHMS 05 39 00 \DECDM +27 50
\SIZE 180 \TYPE S
\FLUX1GHZ 65 \ALPHA varies
\NAMES S147

\RADIO Large faint shell, with spectral break.
\OPTICAL Wispy ring.
\XRAY Possible detection.
\POINT Pulsar within boundary, with faint wind nebula.
\DISTANCE Optical absorption towards stars indicates $>0.36$ and $<0.88$~kpc.
\DATE 19 Feb 2009

\LONGITUDE 182.4 \LATITUDE +4.3
\RAHMS 06 08 10 \DECDM +29 00
\SIZE 50 \TYPE S
\FLUX1GHZ 1.2 \ALPHA 0.4
\NAMES

\RADIO Incomplete shell.
\DATE 2 Apr 2006

\LONGITUDE 184.6 \LATITUDE -5.8
\RAHMS 05 34 31 \DECDM +22 01
\SIZE 7\X/5 \TYPE F
\FLUX1GHZ 1040 \ALPHA 0.30
\NAMES Crab Nebula, 3C144, SN1054

\NOTES This is the remnant of the SN of \AD/1054.
\RADIO Filled-centre, central pulsar, with faint `jet' (or tube) extending from the N edge.
\OPTICAL Strongly polarised filaments, diffuse synchrotron emission, with `jet' faintly visible.
\XRAY Central `torus' around the pulsar.
\POINT Pulsar powering the remnant.
\DISTANCE Proper motions and radial velocities give 2~kpc.
\DATE 27 Mar 2009

\LONGITUDE 189.1 \LATITUDE +3.0
\RAHMS 06 17 00 \DECDM +22 34
\SIZE 45 \TYPE C
\FLUX1GHZ 160 \ALPHA 0.36
\NAMES IC443, 3C157

\RADIO Limb-brightened to NE, with faint extension to the E.
\OPTICAL Brightest to the NE, with faint filaments outside the NE boundary.
\XRAY Shell, brightest to the NE, plus compact source with nebula.
\POINT X-ray source and nebula in S.
\DISTANCE Mean optical velocity suggests 0.7--1.5~kpc, association with S249 gives 1.5--2~kpc.
\DATE 19 Feb 2009

\LONGITUDE 192.8 \LATITUDE -1.1
\RAHMS 06 09 20 \DECDM +17 20
\SIZE 78 \TYPE S
\FLUX1GHZ 20? \ALPHA 0.6?
\NAMES PKS 0607$+$17

\NOTES Has been called G193.3$-$1.5. Has been regarded as part of the Origem Loop, a supposed larger remnant.
\RADIO In complex region.
\OPTICAL Encompasses S261 and S254--258.
\DATE 31 Dec 2001

\LONGITUDE 205.5 \LATITUDE +0.5
\RAHMS 06 39 00 \DECDM +06 30
\SIZE 220 \TYPE S
\FLUX1GHZ 160 \ALPHA 0.5
\NAMES Monoceros Nebula

\RADIO In complex region, parts may be H\II/ regions.
\OPTICAL Large ring, near Rosette nebula.
\XRAY Possibly detected.
\DISTANCE Mean optical velocity suggests 0.8~kpc, low frequency radio absorption suggests 1.6~kpc.
\DATE 19 Feb 2009

\LONGITUDE 206.9 \LATITUDE +2.3
\RAHMS 06 48 40 \DECDM +06 26
\SIZE 60\X/40 \TYPE S?
\FLUX1GHZ 6 \ALPHA 0.5
\NAMES PKS 0646$+$06

\RADIO Diffuse source near the Monoceros Nebula.
\OPTICAL Filaments detected.
\XRAY Possibly detected.
\DATE 2 Apr 2006

\LONGITUDE 260.4 \LATITUDE -3.4
\RAHMS 08 22 10 \DECDM -43 00
\SIZE 60\X/50 \TYPE S
\FLUX1GHZ 130 \ALPHA 0.5
\NAMES Puppis A, MSH 08$-$4{\sl 4}

\NOTES This remnant overlaps the Vela SNR (G263.9$-$3.3).
\RADIO Angular shell, brightest to the E, poorly defined to the W.
\OPTICAL Nebulosity and wisps.
\XRAY Brightest to the E.
\POINT Central possible pulsating X-ray source.
\DISTANCE Association with H\I/ gives 2.2~kpc.
\DATE 19 Feb 2009

\LONGITUDE 261.9 \LATITUDE +5.5
\RAHMS 09 04 20 \DECDM -38 42
\SIZE 40\X/30 \TYPE S
\FLUX1GHZ 10? \ALPHA 0.4?
\NAMES

\RADIO Faint shell with little limb brightening.
\DATE 13 Aug 1998

\LONGITUDE 263.9 \LATITUDE -3.3
\RAHMS 08 34 00 \DECDM -45 50
\SIZE 255 \TYPE C
\FLUX1GHZ 1750 \ALPHA varies
\NAMES Vela (XYZ)

\NOTES This refers to the whole Vela XYZ complex, of which X has at times been classified as a separate (filled-centre) remnant. This remnant is overlapped by G260.4$-$3.4 and G266.2$-$1.2.
\RADIO Large shell, with flatter spectrum component (Vela X), and pulsar nebula.
\OPTICAL Filaments.
\XRAY Patchy shell, with extensions, central nebula and pulsar.
\POINT Pulsar within Vela X, with one-sided `jet'.
\DISTANCE Vela pulsar parallax gives 0.3~kpc, optical spectra and H\I/ studies suggest 0.25~kpc.
\DATE 19 Feb 2009

\LONGITUDE 266.2 \LATITUDE -1.2
\RAHMS 08 52 00 \DECDM -46 20
\SIZE 120 \TYPE S
\FLUX1GHZ 50? \ALPHA 0.3?
\NAMES RX J0852.0$-$4622

\NOTES This remnant overlaps the Vela SNR (G263.9$-$3.3).
\RADIO Incomplete shell, confused by the Vela SNR.
\OPTICAL Nebulosity offset to NE.
\XRAY Non-thermal shell, confused by the Vela SNR, with central source, and possible associated pulsar.
\POINT Central X-ray source, with optical nebula, and possible associated pulsar.
\DISTANCE X-ray data suggest an upper limit of 1~kpc.
\DATE 19 Feb 2009

\LONGITUDE 272.2 \LATITUDE -3.2
\RAHMS 09 06 50 \DECDM -52 07
\SIZE 15? \TYPE S?
\FLUX1GHZ 0.4 \ALPHA 0.6
\NAMES

\RADIO Diffuse shell.
\XRAY Centrally brightened.
\OPTICAL Detected.
\DATE 7 Jan 2004

\LONGITUDE 279.0 \LATITUDE +1.1
\RAHMS 09 57 40 \DECDM -53 15
\SIZE 95 \TYPE S
\FLUX1GHZ 30? \ALPHA 0.6?
\NAMES

\RADIO Faint, incomplete shell.
\POINT Pulsar nearby.
\DATE 13 Aug 1998

\LONGITUDE 284.3 \LATITUDE -1.8
\RAHMS 10 18 15 \DECDM -59 00
\SIZE 24? \TYPE S
\FLUX1GHZ 11? \ALPHA 0.3?
\NAMES MSH 10$-$5{\sl 3}

\NOTES Has been called G284.2$-$1.8.
\RADIO Incomplete, poorly defined shell.
\POINT Pulsar with wind nebula nearby.
\DATE 2 Apr 2006

\LONGITUDE 286.5 \LATITUDE -1.2
\RAHMS 10 35 40 \DECDM -59 42
\SIZE 26\X/6 \TYPE S?
\FLUX1GHZ 1.4? \ALPHA ?
\NAMES

\RADIO Double, elongated arc.
\DATE 13 Aug 1998

\LONGITUDE 289.7 \LATITUDE -0.3
\RAHMS 11 01 15 \DECDM -60 18
\SIZE 18\X/14 \TYPE S
\FLUX1GHZ 6.2 \ALPHA 0.2?
\NAMES

\RADIO Incomplete shell.
\POINT Compact radio source near centre.
\DATE 21 Aug 1996

\LONGITUDE 290.1 \LATITUDE -0.8
\RAHMS 11 03 05 \DECDM -60 56
\SIZE 19\X/14 \TYPE S
\FLUX1GHZ 42 \ALPHA 0.4
\NAMES MSH 11$-$6{\sl 1}A

\RADIO Elongated, clumpy shell.
\OPTICAL Filaments detected.
\XRAY Centrally brightened.
\POINT Pulsar nearby.
\DISTANCE H\I/ absorption indicates $7\PM/1$~kpc.
\DATE 30 Mar 2009

\LONGITUDE 291.0 \LATITUDE -0.1
\RAHMS 11 11 54 \DECDM -60 38
\SIZE 15\X/13 \TYPE C
\FLUX1GHZ 16 \ALPHA 0.29
\NAMES (MSH 11$-$6{\sl 2})

\RADIO Centrally brightened core, with surrounding arcs.
\XRAY Centrally brightened.
\POINT Central compact X-ray source.
\DATE 13 Aug 1998

\LONGITUDE 292.0 \LATITUDE +1.8
\RAHMS 11 24 36 \DECDM -59 16
\SIZE 12\X/8 \TYPE C
\FLUX1GHZ 15 \ALPHA 0.4
\NAMES MSH 11$-$5{\sl 4}

\RADIO Centrally brightened source surrounded by a plateau of faint emission.
\OPTICAL Oxygen rich.
\XRAY Ring of emission, with diffuse central nebula and pulsar.
\POINT Central pulsar.
\DISTANCE H\I/ absorption implies 6.0~kpc.
\DATE 19 Feb 2009

\LONGITUDE 292.2 \LATITUDE -0.5
\RAHMS 11 19 20 \DECDM -61 28
\SIZE 20\X/15 \TYPE S
\FLUX1GHZ 7 \ALPHA 0.5
\NAMES

\RADIO Shell.
\XRAY Detected.
\POINT Central, young pulsar with X-ray nebula.
\DISTANCE H\I/ absorption indicates 8.4~kpc.
\DATE 19 Feb 2009

\LONGITUDE 293.8 \LATITUDE +0.6
\RAHMS 11 35 00 \DECDM -60 54
\SIZE 20 \TYPE C
\FLUX1GHZ 5? \ALPHA 0.6?
\NAMES

\RADIO Central source, with faint extended plateau.
\DATE 21 Aug 1996

\LONGITUDE 294.1 \LATITUDE -0.0
\RAHMS 11 36 10 \DECDM -61 38
\SIZE 40 \TYPE S
\FLUX1GHZ $>$2? \ALPHA ?
\NAMES

\RADIO Faint shell.
\DATE 21 Aug 1996

\LONGITUDE 296.1 \LATITUDE -0.5
\RAHMS 11 51 10 \DECDM -62 34
\SIZE 37\X/25 \TYPE S
\FLUX1GHZ 8? \ALPHA 0.6?
\NAMES

\NOTES Incorporates the previously catalogued remnant G296.1$-$0.7. Has been called G296.05$-$0.50.
\RADIO Irregular shell, with nearby H\II/ regions.
\OPTICAL Detected.
\XRAY Detected.
\DATE 19 Aug 1998

\LONGITUDE 296.5 \LATITUDE +10.0
\RAHMS 12 09 40 \DECDM -52 25
\SIZE 90\X/65 \TYPE S
\FLUX1GHZ 48 \ALPHA 0.5
\NAMES PKS 1209$-$51/52

\NOTES Has been called G296.5$+$9.7.
\RADIO Shell with two bright limbs.
\OPTICAL Detected.
\XRAY Incomplete shell, with central pulsar.
\POINT Central pulsar.
\DATE 19 Feb 2009

\LONGITUDE 296.8 \LATITUDE -0.3
\RAHMS 11 58 30 \DECDM -62 35
\SIZE 20\X/14 \TYPE S
\FLUX1GHZ 9 \ALPHA 0.6
\NAMES 1156$-$62

\RADIO Shell, brighter to the NW.
\DISTANCE H\I/ absorption gives 9.6~kpc.
\XRAY Detected.
\DATE 26 Apr 2006

\LONGITUDE 298.5 \LATITUDE -0.3
\RAHMS 12 12 40 \DECDM -62 52
\SIZE 5? \TYPE ?
\FLUX1GHZ 5? \ALPHA 0.4?
\NAMES

\RADIO Not well resolved, may be part of a larger ring?
\DATE 16 Mar 2009

\LONGITUDE 298.6 \LATITUDE -0.0
\RAHMS 12 13 41 \DECDM -62 37
\SIZE 12\X/9 \TYPE S
\FLUX1GHZ 5? \ALPHA 0.3
\NAMES

\NOTES Has been called G298.6$-$0.1.
\RADIO Incomplete shell, in complex region.
\DATE 16 Mar 2009

\LONGITUDE 299.2 \LATITUDE -2.9
\RAHMS 12 15 13 \DECDM -65 30
\SIZE 18\X/11 \TYPE S
\FLUX1GHZ 0.5? \ALPHA ?
\NAMES

\RADIO Faint source.
\XRAY Centrally brightened with shell at higher energies.
\OPTICAL Filaments in W.
\DATE 19 Feb 2009

\LONGITUDE 299.6 \LATITUDE -0.5
\RAHMS 12 21 45 \DECDM -63 09
\SIZE 13 \TYPE S
\FLUX1GHZ 1.0? \ALPHA ?
\NAMES

\RADIO Faint shell, brightest to E.
\DATE 21 Aug 1996

\LONGITUDE 301.4 \LATITUDE -1.0
\RAHMS 12 37 55 \DECDM -63 49
\SIZE 37\X/23 \TYPE S
\FLUX1GHZ 2.1? \ALPHA ?
\NAMES

\RADIO Faint, incomplete shell, with possible extension to southwest.
\DATE 21 Aug 1996

\LONGITUDE 302.3 \LATITUDE +0.7
\RAHMS 12 45 55 \DECDM -62 08
\SIZE 17 \TYPE S
\FLUX1GHZ 5? \ALPHA 0.4?
\NAMES

\RADIO Distorted shell, in complex region, with possibly associated filament.
\DATE 21 Aug 1996

\LONGITUDE 304.6 \LATITUDE +0.1
\RAHMS 13 05 59 \DECDM -62 42
\SIZE 8 \TYPE S
\FLUX1GHZ 14 \ALPHA 0.5
\NAMES Kes 17

\RADIO Incomplete shell.
\DISTANCE Possible limit of $>9.7$~kpc from H\I/ absorption.
\DATE 16 Mar 2009

\LONGITUDE 308.1 \LATITUDE -0.7
\RAHMS 13 37 37 \DECDM -63 04
\SIZE 13 \TYPE S
\FLUX1GHZ 1.2? \ALPHA ?
\NAMES

\RADIO Faint shell.
\DATE 21 Aug 1996

\LONGITUDE 308.8 \LATITUDE -0.1
\RAHMS 13 42 30 \DECDM -62 23
\SIZE 30\X/20? \TYPE C?
\FLUX1GHZ 15? \ALPHA 0.4?
\NAMES

\NOTES Incorporates previous catalogued remnant G308.7$+$0.0.
\RADIO Bright ridge in north, and arc to south.
\POINT Pulsar near centre of remnant.
\DATE 16 Mar 2009

\LONGITUDE 309.2 \LATITUDE -0.6
\RAHMS 13 46 31 \DECDM -62 54
\SIZE 15\X/12 \TYPE S
\FLUX1GHZ 7? \ALPHA 0.4?
\NAMES

\NOTES Has been called G309.2$-$0.7.
\RADIO Distorted shell.
\XRAY Extended emission, with unrelated central source.
\DATE 27 Mar 2009

\LONGITUDE 309.8 \LATITUDE +0.0
\RAHMS 13 50 30 \DECDM -62 05
\SIZE 25\X/19 \TYPE S
\FLUX1GHZ 17 \ALPHA 0.5
\NAMES

\RADIO Distorted shell.
\POINT Steep radio spectrum source near the centre of the remnant.
\DATE 13 Aug 1998

\LONGITUDE 310.6 \LATITUDE -0.3
\RAHMS 13 58 00 \DECDM -62 09
\SIZE 8 \TYPE S
\FLUX1GHZ 5? \ALPHA ?
\NAMES Kes 20B

\RADIO Asymmetric shell.
\DATE 16 Mar 2009

\LONGITUDE 310.8 \LATITUDE -0.4
\RAHMS 14 00 00 \DECDM -62 17
\SIZE 12 \TYPE S
\FLUX1GHZ 6? \ALPHA ?
\NAMES Kes 20A

\RADIO Arc in E, in complex region.
\DATE 16 Mar 2009

\LONGITUDE 311.5 \LATITUDE -0.3
\RAHMS 14 05 38 \DECDM -61 58
\SIZE 5 \TYPE S
\FLUX1GHZ 3? \ALPHA 0.5
\NAMES

\RADIO Shell, not well resolved.
\DATE 16 Mar 2009

\LONGITUDE 312.4 \LATITUDE -0.4
\RAHMS 14 13 00 \DECDM -61 44
\SIZE 38 \TYPE S
\FLUX1GHZ 45 \ALPHA 0.36
\NAMES

\RADIO Irregular, incomplete shell.
\POINT Nearby $\gamma$-ray sources and pulsars.
\XRAY Weak emission in W.
\DISTANCE H\I/ absorption suggests $>6$~kpc and possibly $>14$~kpc.
\DATE 11 Jan 2004

\LONGITUDE 312.5 \LATITUDE -3.0
\RAHMS 14 21 00 \DECDM -64 12
\SIZE 20\X/18 \TYPE S
\FLUX1GHZ 3.5? \ALPHA ?
\NAMES

\RADIO Distorted shell.
\DATE 11 Jan 2004

\LONGITUDE 315.1 \LATITUDE +2.7
\RAHMS 14 24 30 \DECDM -57 50
\SIZE 190\X/150 \TYPE S
\FLUX1GHZ ? \ALPHA ?
\NAMES

\RADIO Poorly defined shell.
\OPTICAL Filaments, brighter in NE.
\DATE 16 Mar 2009

\LONGITUDE 315.4 \LATITUDE -2.3
\RAHMS 14 43 00 \DECDM -62 30
\SIZE 42 \TYPE S
\FLUX1GHZ 49 \ALPHA 0.6
\NAMES RCW 86, MSH 14$-$6{\sl 3}

\NOTES Possibly the remnant of the SN of \AD/185?
\RADIO Shell, brightest to the SW.
\OPTICAL Bright, radiative filaments, with some faint Balmer dominated filaments.
\XRAY Partial shell, with thermal and non-thermal emission.
\POINT Several X-ray sources.
\DISTANCE Optical observations imply 2.3~kpc.
\DATE 16 Mar 2009

\LONGITUDE 315.4 \LATITUDE -0.3
\RAHMS 14 35 55 \DECDM -60 36
\SIZE 24\X/13 \TYPE ?
\FLUX1GHZ 8 \ALPHA 0.4
\NAMES

\RADIO Irregular non-thermal emission, with H\II/ region superposed in E.
\DATE 16 Mar 2009

\LONGITUDE 315.9 \LATITUDE -0.0
\RAHMS 14 38 25 \DECDM -60 11
\SIZE 25\X/14 \TYPE S
\FLUX1GHZ 0.8? \ALPHA ?
\NAMES

\NOTES Has been called G315.8$-$0.0.
\RADIO Faint, distorted shell, with steep-spectrum `jet'?
\DATE 14 Aug 1998

\LONGITUDE 316.3 \LATITUDE -0.0
\RAHMS 14 41 30 \DECDM -60 00
\SIZE 29\X/14 \TYPE S
\FLUX1GHZ 20? \ALPHA 0.4
\NAMES (MSH 14$-$5{\sl 7})

\RADIO Distorted shell, with possible `blowout'.
\XRAY Detected.
\DISTANCE H\I/ absorption data suggests $>7.2$~kpc.
\DATE 10 Oct 2001

\LONGITUDE 317.3 \LATITUDE -0.2
\RAHMS 14 49 40 \DECDM -59 46
\SIZE 11 \TYPE S
\FLUX1GHZ 4.7? \ALPHA ?
\NAMES

\RADIO Incomplete shell.
\DATE 21 Aug 1996

\LONGITUDE 318.2 \LATITUDE +0.1
\RAHMS 14 54 50 \DECDM -59 04
\SIZE 40\X/35 \TYPE S
\FLUX1GHZ $>$3.9? \ALPHA ?
\NAMES

\RADIO Faint shell, with central H\II/ region.
\XRAY Sources within remnant.
\DATE 10 Oct 2001

\LONGITUDE 318.9 \LATITUDE +0.4
\RAHMS 14 58 30 \DECDM -58 29
\SIZE 30\X/14 \TYPE C
\FLUX1GHZ 4? \ALPHA 0.2?
\NAMES

\NOTES May not be a SNR?
\RADIO Complex arcs, with off-centre core.
\DATE 13 Aug 1998

\LONGITUDE 320.4 \LATITUDE -1.2
\RAHMS 15 14 30 \DECDM -59 08
\SIZE 35 \TYPE C
\FLUX1GHZ 60? \ALPHA 0.4
\NAMES MSH 15$-$5{\sl 2}, RCW 89

\NOTES Has been suggested as the remnant of the SN of \AD/185?
\RADIO Ragged shell.
\OPTICAL RCW 89 is the \Halpha/ emitting region to the NW.
\XRAY Partial shell, central nebula and pulsar and `jet'.
\POINT Radio and X-ray pulsar, with wind nebula.
\DISTANCE H\I/ absorption indicates 5.2~kpc.
\DATE 19 Feb 2009

\LONGITUDE 320.6 \LATITUDE -1.6
\RAHMS 15 17 50 \DECDM -59 16
\SIZE 60\X/30 \TYPE S
\FLUX1GHZ ? \ALPHA ?
\NAMES

\RADIO Faint shell, overlapping G320.4$-$1.2 in W.
\DATE 19 Aug 1998

\LONGITUDE 321.9 \LATITUDE -1.1
\RAHMS 15 23 45 \DECDM -58 13
\SIZE 28 \TYPE S
\FLUX1GHZ $>$3.4? \ALPHA ?
\NAMES

\RADIO Faint shell.
\DATE 7 Jan 2004

\LONGITUDE 321.9 \LATITUDE -0.3
\RAHMS 15 20 40 \DECDM -57 34
\SIZE 31\X/23 \TYPE S
\FLUX1GHZ 13 \ALPHA 0.3
\NAMES

\RADIO Shell brighter to the W, with Cir X-1 to N.
\POINT Compact, probably thermal source at S edge.
\DATE 19 Feb 2009

\LONGITUDE 322.5 \LATITUDE -0.1
\RAHMS 15 23 23 \DECDM -57 06
\SIZE 15 \TYPE C
\FLUX1GHZ 1.5 \ALPHA 0.4
\NAMES

\RADIO Shell with central extended source.
\POINT PN Pe 2-8 within boundary.
\DATE 13 Aug 1998

\LONGITUDE 323.5 \LATITUDE +0.1
\RAHMS 15 28 42 \DECDM -56 21
\SIZE 13 \TYPE S
\FLUX1GHZ 3? \ALPHA 0.4?
\NAMES

\RADIO Distorted shell, confused with thermal emission.
\POINT Compact, probably thermal source near centre.
\DATE 16 Mar 2009

\LONGITUDE 326.3 \LATITUDE -1.8
\RAHMS 15 53 00 \DECDM -56 10
\SIZE 38 \TYPE C
\FLUX1GHZ 145 \ALPHA varies
\NAMES MSH 15$-$5{\sl 6}

\NOTES Has been called G326.2$-$1.7.
\RADIO Shell, with elongated, flat-spectrum core.
\OPTICAL Emission around the shell.
\XRAY Shell, with central extended emission.
\DATE 10 Oct 2001

\LONGITUDE 327.1 \LATITUDE -1.1
\RAHMS 15 54 25 \DECDM -55 09
\SIZE 18 \TYPE C
\FLUX1GHZ 7? \ALPHA ?
\NAMES

\RADIO Shell, with off-centre core.
\XRAY Diffuse, with core.
\DATE 11 Jan 2004

\LONGITUDE 327.2 \LATITUDE -0.1
\RAHMS 15 50 55 \DECDM -54 18
\SIZE 5 \TYPE S
\FLUX1GHZ 0.4 \ALPHA ?
\NAMES

\NOTES Has been called G327.24$-$0.13.
\RADIO Shell, possibly with central emission.
\POINT Central pulsar (magnetar).
\DATE 16 Mar 2009

\LONGITUDE 327.4 \LATITUDE +0.4
\RAHMS 15 48 20 \DECDM -53 49
\SIZE 21 \TYPE S
\FLUX1GHZ 30? \ALPHA 0.6
\NAMES Kes 27

\NOTES Has been called G327.3$+$0.4 and G327.3$+$0.5.
\RADIO Incomplete, multi-arc shell, brightest to the SE.
\XRAY Diffuse, best defined to E.
\DISTANCE H\I/ absorption indicates 4.3 to 5.4~kpc.
\DATE 16 Mar 2009

\LONGITUDE 327.4 \LATITUDE +1.0
\RAHMS 15 46 48 \DECDM -53 20
\SIZE 14 \TYPE S
\FLUX1GHZ 1.9? \ALPHA ?
\NAMES

\RADIO Asymmetric shell.
\DATE 10 Oct 2001

\LONGITUDE 327.6 \LATITUDE +14.6
\RAHMS 15 02 50 \DECDM -41 56
\SIZE 30 \TYPE S
\FLUX1GHZ 19 \ALPHA 0.6
\NAMES SN1006, PKS 1459$-$41

\NOTES This is the remnant of the SN of \AD/1006.
\RADIO Shell, with two bright arcs.
\OPTICAL Filaments to the NW, with broad \Halpha/ component.
\XRAY Thermal shell, with non-thermal limb-brightened arcs.
\POINT The background Schweizer--Middleditch star is near the middle of the remnant.
\DISTANCE Optical spectra and proper motion indicate 2.2~kpc.
\DATE 30 Mar 2009

\LONGITUDE 328.4 \LATITUDE +0.2
\RAHMS 15 55 30 \DECDM -53 17
\SIZE 5 \TYPE F
\FLUX1GHZ 15 \ALPHA 0.0
\NAMES (MSH 15$-$5{\sl 7})

\RADIO Amorphous emission, with central bar.
\XRAY Detected at high energies.
\DISTANCE H\I/ absorption indicates $>17.4$~kpc.
\DATE 27 Mar 2009

\LONGITUDE 329.7 \LATITUDE +0.4
\RAHMS 16 01 20 \DECDM -52 18
\SIZE 40\X/33 \TYPE S
\FLUX1GHZ $>$34? \ALPHA ?
\NAMES

\RADIO Diffuse shell, in complex region.
\DATE 16 Mar 2009

\LONGITUDE 330.0 \LATITUDE +15.0
\RAHMS 15 10 00 \DECDM -40 00
\SIZE 180? \TYPE S
\FLUX1GHZ 350? \ALPHA 0.5?
\NAMES Lupus Loop

\RADIO Low surface brightness loop with H\I/ shell.
\XRAY Detected, with central source.
\POINT Central, possibly pulsating, X-ray source.
\DATE 19 Feb 2009

\LONGITUDE 330.2 \LATITUDE +1.0
\RAHMS 16 01 06 \DECDM -51 34
\SIZE 11 \TYPE S?
\FLUX1GHZ 5? \ALPHA 0.3
\NAMES

\RADIO Clumpy non-thermal emission, possibly a distorted shell.
\XRAY Shell.
\DISTANCE H\I/ absorption indicates $>4.9$~kpc.
\DATE 26 Apr 2006

\LONGITUDE 332.0 \LATITUDE +0.2
\RAHMS 16 13 17 \DECDM -50 53
\SIZE 12 \TYPE S
\FLUX1GHZ 8? \ALPHA 0.5
\NAMES

\RADIO Incomplete shell.
\DATE 17 Oct 2001

\LONGITUDE 332.4 \LATITUDE -0.4
\RAHMS 16 17 33 \DECDM -51 02
\SIZE 10 \TYPE S
\FLUX1GHZ 28 \ALPHA 0.5
\NAMES RCW 103

\RADIO Shell, brightest to the S.
\OPTICAL Filaments correspond well to the radio shell, brightest in SE.
\XRAY Brightest to NW, with point source near centre.
\POINT Central, variable X-ray source, and nearby pulsar.
\DISTANCE H\I/ absorption indicates 3.1~kpc.
\DATE 16 Mar 2009

\LONGITUDE 332.4 \LATITUDE +0.1
\RAHMS 16 15 20 \DECDM -50 42
\SIZE 15 \TYPE S
\FLUX1GHZ 26 \ALPHA 0.5
\NAMES MSH 16$-$5{\sl 1}, Kes 32

\NOTES Has been called G332.4$+$0.2.
\RADIO Distorted shell, with thermal jet and plume adjacent.
\XRAY Shell, brightest to NW.
\POINT Pulsar nearby.
\DATE 16 Mar 2009

\LONGITUDE 332.5 \LATITUDE -5.6
\RAHMS 16 43 20 \DECDM -54 30
\SIZE 35 \TYPE S
\FLUX1GHZ 2? \ALPHA 0.7?
\NAMES

\RADIO Bipolar shell, with central emission also.
\OPTICAL Patchy filaments.
\XRAY Emission from centre.
\DATE 16 Mar 2009

\LONGITUDE 335.2 \LATITUDE +0.1
\RAHMS 16 27 45 \DECDM -48 47
\SIZE 21 \TYPE S
\FLUX1GHZ 16 \ALPHA 0.5
\NAMES

\RADIO Well defined shell.
\POINT Old pulsar within remnant boundary.
\DATE 16 Mar 2009

\LONGITUDE 336.7 \LATITUDE +0.5
\RAHMS 16 32 11 \DECDM -47 19
\SIZE 14\X/10 \TYPE S
\FLUX1GHZ 6 \ALPHA 0.5
\NAMES

\RADIO Irregular shell.
\DATE 13 Aug 1998

\LONGITUDE 337.0 \LATITUDE -0.1
\RAHMS 16 35 57 \DECDM -47 36
\SIZE 1.5 \TYPE S
\FLUX1GHZ 1.5 \ALPHA 0.6?
\NAMES (CTB 33)

\NOTES This entry refers to a small ($1\fmin5$) SNR, not the larger previously catalogued G337.0$-$0.1.
\RADIO Shell, in a complex region.
\DISTANCE Association with CTB 33 gives 11~kpc.
\POINT Associated with a soft gamma repeater.
\DATE 17 Oct 2001

\LONGITUDE 337.2 \LATITUDE -0.7
\RAHMS 16 39 28 \DECDM -47 51
\SIZE 6 \TYPE S
\FLUX1GHZ 1.5 \ALPHA 0.4
\NAMES

\RADIO Shell, brighter in S.
\XRAY Extended emission.
\DISTANCE H\I/ absorption suggests 2.0 to 9.3~kpc.
\DATE 16 Mar 2009

\LONGITUDE 337.2 \LATITUDE +0.1
\RAHMS 16 35 55 \DECDM -47 20
\SIZE 3\X/2 \TYPE ?
\FLUX1GHZ 1.5? \ALPHA ?
\NAMES

\RADIO Not well defined.
\XRAY Detected.
\DISTANCE Association with H\I/ hole gives 14~kpc.
\DATE 19 Feb 2009

\LONGITUDE 337.3 \LATITUDE +1.0
\RAHMS 16 32 39 \DECDM -46 36
\SIZE 15\X/12 \TYPE S
\FLUX1GHZ 16 \ALPHA 0.55
\NAMES Kes 40

\RADIO Nearly complete shell.
\DATE 13 Aug 1998

\LONGITUDE 337.8 \LATITUDE -0.1
\RAHMS 16 39 01 \DECDM -46 59
\SIZE 9\X/6 \TYPE S
\FLUX1GHZ 18 \ALPHA 0.5
\NAMES Kes 41

\RADIO Distorted shell.
\XRAY Centrally brightened.
\DISTANCE H\I/ absorption suggests 11~kpc.
\DATE 16 Mar 2009

\LONGITUDE 338.1 \LATITUDE +0.4
\RAHMS 16 37 59 \DECDM -46 24
\SIZE 15? \TYPE S
\FLUX1GHZ 4? \ALPHA 0.4
\NAMES

\RADIO Arc in NE, merging with thermal emission in S.
\OPTICAL Detected.
\XRAY Detected.
\DATE 31 Mar 2006

\LONGITUDE 338.3 \LATITUDE -0.0
\RAHMS 16 41 00 \DECDM -46 34
\SIZE 8 \TYPE C?
\FLUX1GHZ 7? \ALPHA ?
\NAMES

\RADIO Irregular shell, in complex region.
\XRAY Central X-ray source and nebula.
\DISTANCE H\I/ observations suggest 11~kpc.
\DATE 1 Apr 2006

\LONGITUDE 338.5 \LATITUDE +0.1
\RAHMS 16 41 09 \DECDM -46 19
\SIZE 9 \TYPE ?
\FLUX1GHZ 12? \ALPHA ?
\NAMES

\RADIO Circle of non-thermal emission in complex region, not well defined.
\DISTANCE H\I/ absorption suggests 11~kpc.
\DATE 16 Mar 2009

\LONGITUDE 340.4 \LATITUDE +0.4
\RAHMS 16 46 31 \DECDM -44 39
\SIZE 10\X/7 \TYPE S
\FLUX1GHZ 5 \ALPHA 0.4
\NAMES

\RADIO Distorted shell, elongated east--west.
\DATE 13 Aug 1998

\LONGITUDE 340.6 \LATITUDE +0.3
\RAHMS 16 47 41 \DECDM -44 34
\SIZE 6 \TYPE S
\FLUX1GHZ 5? \ALPHA 0.4?
\NAMES

\RADIO Incomplete shell.
\OPTICAL Possible associated filaments.
\DISTANCE H\I/ absorption suggests 15~kpc.
\DATE 16 Mar 2009

\LONGITUDE 341.2 \LATITUDE +0.9
\RAHMS 16 47 35 \DECDM -43 47
\SIZE 22\X/16 \TYPE C
\FLUX1GHZ 1.5? \ALPHA 0.6?
\NAMES

\RADIO Incomplete shell, with extension to SW.
\POINT Pulsar in W, with wind nebula.
\DATE 8 Nov 2001

\LONGITUDE 341.9 \LATITUDE -0.3
\RAHMS 16 55 01 \DECDM -44 01
\SIZE 7 \TYPE S
\FLUX1GHZ 2.5 \ALPHA 0.5
\NAMES

\RADIO Incomplete shell, brightest to NE.
\DATE 27 Jan 2004

\LONGITUDE 342.0 \LATITUDE -0.2
\RAHMS 16 54 50 \DECDM -43 53
\SIZE 12\X/9 \TYPE S
\FLUX1GHZ 3.5? \ALPHA 0.4?
\NAMES

\RADIO Distorted shell.
\DATE 1 Aug 2000

\LONGITUDE 342.1 \LATITUDE +0.9
\RAHMS 16 50 43 \DECDM -43 04
\SIZE 10\X/9 \TYPE S
\FLUX1GHZ 0.5? \ALPHA ?
\NAMES

\RADIO Incomplete shell.
\DATE 13 Aug 1998

\LONGITUDE 343.0 \LATITUDE -6.0
\RAHMS 17 25 00 \DECDM -46 30
\SIZE 250 \TYPE S
\FLUX1GHZ ? \ALPHA ?
\NAMES RCW 114

\RADIO Faint, poorly defined.
\OPTICAL Filamentary shell.
\DATE 18 Feb 2009

\LONGITUDE 343.1 \LATITUDE -2.3
\RAHMS 17 08 00 \DECDM -44 16
\SIZE 32? \TYPE C?
\FLUX1GHZ 8? \ALPHA 0.5?
\NAMES

\RADIO Incomplete shell?
\XRAY Pulsar wind nebula.
\POINT Pulsar near edge, with wind nebula.
\DATE 27 Jan 2004

\LONGITUDE 343.1 \LATITUDE -0.7
\RAHMS 17 00 25 \DECDM -43 14
\SIZE 27\X/21 \TYPE S
\FLUX1GHZ 7.8 \ALPHA 0.55
\NAMES

\RADIO Shell, with smaller thermal shell adjacent.
\DATE 1 Aug 2000

\LONGITUDE 344.7 \LATITUDE -0.1
\RAHMS 17 03 51 \DECDM -41 42
\SIZE 10 \TYPE C?
\FLUX1GHZ 2.5? \ALPHA 0.5
\NAMES

\RADIO Asymmetric shell, with possible core.
\XRAY Detected.
\DATE 16 Mar 2009

\LONGITUDE 345.7 \LATITUDE -0.2
\RAHMS 17 07 20 \DECDM -40 53
\SIZE 6 \TYPE S
\FLUX1GHZ 0.6? \ALPHA ?
\NAMES

\RADIO Poorly defined diffuse shell.
\POINT Old pulsar nearby.
\DATE 13 Aug 1998

\LONGITUDE 346.6 \LATITUDE -0.2
\RAHMS 17 10 19 \DECDM -40 11
\SIZE 8 \TYPE S
\FLUX1GHZ 8? \ALPHA 0.5?
\NAMES

\RADIO Irregular shell.
\DATE 16 Mar 2009

\LONGITUDE 347.3 \LATITUDE -0.5
\RAHMS 17 13 50 \DECDM -39 45
\SIZE 65\X/55 \TYPE S?
\FLUX1GHZ ? \ALPHA ?
\NAMES

\RADIO Faint emission.
\XRAY Non-thermal, limb-brightened to W, with central source.
\POINT Central X-ray source.
\DISTANCE Association with molecular clouds and X-ray observations imply 1.3~kpc.
\DATE 19 Feb 2009

\LONGITUDE 348.5 \LATITUDE -0.0
\RAHMS 17 15 26 \DECDM -38 28
\SIZE 10? \TYPE S?
\FLUX1GHZ 10? \ALPHA 0.4?
\NAMES

\RADIO Arc, overlapping G348.5$+$0.1.
\DATE 16 Mar 2009

\LONGITUDE 348.5 \LATITUDE +0.1
\RAHMS 17 14 06 \DECDM -38 32
\SIZE 15 \TYPE S
\FLUX1GHZ 72 \ALPHA 0.3
\NAMES CTB 37A

\RADIO Shell, poorly define to S and W, overlapping G348.5$-$0.0 in E.
\DISTANCE H\I/ absorption indicates 8.0~kpc.
\DATE 16 Mar 2009

\LONGITUDE 348.7 \LATITUDE +0.3
\RAHMS 17 13 55 \DECDM -38 11
\SIZE 17? \TYPE S
\FLUX1GHZ 26 \ALPHA 0.3
\NAMES CTB 37B

\RADIO Incomplete shell with faint eastern extensions.
\XRAY Diffuse emission.
\DISTANCE H\I/ absorption indicates 8.0~kpc.
\DATE 19 Feb 2009

\LONGITUDE 349.2 \LATITUDE -0.1
\RAHMS 17 17 15 \DECDM -38 04
\SIZE 9\X/6 \TYPE S
\FLUX1GHZ 1.4? \ALPHA ?
\NAMES

\RADIO Elongated shell, adjacent to bright H\II/ region.
\DATE 27 Aug 1996

\LONGITUDE 349.7 \LATITUDE +0.2
\RAHMS 17 17 59 \DECDM -37 26
\SIZE 2.5\X/2 \TYPE S
\FLUX1GHZ 20 \ALPHA 0.5
\NAMES

\RADIO Incomplete clumpy shell, with enhancement to the S.
\DISTANCE H\I/ absorption indicates 14.8~kpc, association with OH features gives 22~kpc.
\XRAY Irregular shell, brighter to S and E.
\DATE 16 Mar 2009

\LONGITUDE 350.0 \LATITUDE -2.0
\RAHMS 17 27 50 \DECDM -38 32
\SIZE 45 \TYPE S
\FLUX1GHZ 26 \ALPHA 0.4
\NAMES

\NOTES Incorporates the previously catalogued G350.0$-$1.8 in the NW.
\RADIO Shell, brightest in NW.
\DATE 13 Aug 1998

\LONGITUDE 350.1 \LATITUDE -0.3
\RAHMS 17 17 40 \DECDM -37 24
\SIZE 4? \TYPE ?
\FLUX1GHZ 6? \ALPHA 0.8?
\NAMES

\RADIO Several clumps of emission.
\XRAY Diffuse emission, with compact source.
\DISTANCE H\I/ absorption indicates 4.5 to 10.7~kpc, possible interaction with molecular cloud indicates 4.5~kpc.
\POINT X-ray source.
\DATE 27 Mar 2009

\LONGITUDE 351.2 \LATITUDE +0.1
\RAHMS 17 22 27 \DECDM -36 11
\SIZE 7 \TYPE C?
\FLUX1GHZ 5? \ALPHA 0.4
\NAMES

\NOTES Has been called G351.3$+$0.2.
\RADIO Distorted shell, with possible flat-spectrum core.
\DATE 13 Aug 1998

\LONGITUDE 351.7 \LATITUDE +0.8
\RAHMS 17 21 00 \DECDM -35 27
\SIZE 18\X/14 \TYPE S
\FLUX1GHZ 10 \ALPHA 0.5?
\NAMES

\RADIO Elongated shell, adjacent to bright H\II/ region.
\POINT Pulsar nearby.
\DATE 19 Feb 2009

\LONGITUDE 351.9 \LATITUDE -0.9
\RAHMS 17 28 52 \DECDM -36 16
\SIZE 12\X/9 \TYPE S
\FLUX1GHZ 1.8? \ALPHA ?
\NAMES

\RADIO Asymmetric shell.
\DATE 21 Aug 1996

\LONGITUDE 352.7 \LATITUDE -0.1
\RAHMS 17 27 40 \DECDM -35 07
\SIZE 8\X/6 \TYPE S
\FLUX1GHZ 4 \ALPHA 0.6
\NAMES

\RADIO Distorted shell.
\XRAY Detected.
\DATE 10 Oct 2001

\LONGITUDE 353.6 \LATITUDE -0.7
\RAHMS 17 32 00 \DECDM -34 44
\SIZE 30 \TYPE S
\FLUX1GHZ 2.5? \ALPHA ?
\NAMES

\RADIO Shell, brighter to S.
\XRAY Detected.
\DATE 16 Mar 2009

\LONGITUDE 353.9 \LATITUDE -2.0
\RAHMS 17 38 55 \DECDM -35 11
\SIZE 13 \TYPE S
\FLUX1GHZ 1? \ALPHA 0.5?
\NAMES

\RADIO Shell, with central double source.
\DATE 8 Nov 2001

\LONGITUDE 354.1 \LATITUDE +0.1
\RAHMS 17 30 28 \DECDM -33 46
\SIZE 15\X/3? \TYPE C?
\FLUX1GHZ ? \ALPHA varies
\NAMES

\NOTES Is this a SNR?
\RADIO Elongated N--S.
\POINT Pulsar at S tip.
\DATE 30 Mar 2009

\LONGITUDE 354.8 \LATITUDE -0.8
\RAHMS 17 36 00 \DECDM -33 42
\SIZE 19 \TYPE S
\FLUX1GHZ 2.8? \ALPHA ?
\NAMES

\RADIO Distorted shell.
\DATE 1 Aug 2000

\LONGITUDE 355.4 \LATITUDE +0.7
\RAHMS 17 31 20 \DECDM -32 26
\SIZE 25 \TYPE S
\FLUX1GHZ 5? \ALPHA ?
\NAMES

\RADIO Faint, incomplete shell.
\DATE 16 Mar 2009

\LONGITUDE 355.6 \LATITUDE -0.0
\RAHMS 17 35 16 \DECDM -32 38
\SIZE 8\X/6 \TYPE S
\FLUX1GHZ 3? \ALPHA ?
\NAMES

\RADIO Well defined shell.
\XRAY Detected.
\DATE 19 Feb 2009

\LONGITUDE 355.9 \LATITUDE -2.5
\RAHMS 17 45 53 \DECDM -33 43
\SIZE 13 \TYPE S
\FLUX1GHZ 8 \ALPHA 0.5
\NAMES

\RADIO Distorted shell, brightest to SE.
\DATE 13 Aug 1998

\LONGITUDE 356.2 \LATITUDE +4.5
\RAHMS 17 19 00 \DECDM -29 40
\SIZE 25 \TYPE S
\FLUX1GHZ 4 \ALPHA 0.7
\NAMES

\NOTES Has been called G356.2$+$4.4.
\RADIO Faint shell.
\DATE 1 Apr 2006

\LONGITUDE 356.3 \LATITUDE -0.3
\RAHMS 17 37 56 \DECDM -32 16
\SIZE 11\X/7 \TYPE S
\FLUX1GHZ 3? \ALPHA ?
\NAMES

\RADIO Elongated shell, brighter in N.
\DATE 11 Jan 2004

\LONGITUDE 356.3 \LATITUDE -1.5
\RAHMS 17 42 35 \DECDM -32 52
\SIZE 20\X/15 \TYPE S
\FLUX1GHZ 3? \ALPHA ?
\NAMES

\RADIO Double arc.
\DATE 30 Jun 1995

\LONGITUDE 357.7 \LATITUDE -0.1
\RAHMS 17 40 29 \DECDM -30 58
\SIZE 8\X/3? \TYPE ?
\FLUX1GHZ 37 \ALPHA 0.4
\NAMES MSH 17$-$3{\sl 9}

\NOTES Has been suggested that this is not a SNR.
\RADIO Multiple arcs and filaments, with compact H\II/ region at W edge.
\XRAY Detected.
\DISTANCE H\I/ absorption suggests beyond Galactic Centre.
\DATE 18 Feb 2009

\LONGITUDE 357.7 \LATITUDE +0.3
\RAHMS 17 38 35 \DECDM -30 44
\SIZE 24 \TYPE S
\FLUX1GHZ 10 \ALPHA 0.4?
\NAMES

\RADIO Non-thermal shell in complex region.
\DATE 18 Feb 2009

\LONGITUDE 358.0 \LATITUDE +3.8
\RAHMS 17 26 00 \DECDM -28 36
\SIZE 38 \TYPE S
\FLUX1GHZ 1.5? \ALPHA ?
\NAMES

\RADIO Faint shell.
\DATE 12 Oct 2001

\LONGITUDE 358.1 \LATITUDE +0.1
\RAHMS 17 37 00 \DECDM -29 59
\SIZE 20 \TYPE S
\FLUX1GHZ 2? \ALPHA ?
\NAMES

\RADIO Faint shell.
\DATE 16 Mar 2009

\LONGITUDE 358.5 \LATITUDE -0.9
\RAHMS 17 46 10 \DECDM -30 40
\SIZE 17 \TYPE S
\FLUX1GHZ 4? \ALPHA ?
\NAMES

\RADIO Shell, brighter to NE.
\DATE 16 Mar 2009

\LONGITUDE 359.0 \LATITUDE -0.9
\RAHMS 17 46 50 \DECDM -30 16
\SIZE 23 \TYPE S
\FLUX1GHZ 23 \ALPHA 0.5
\NAMES

\RADIO Incomplete shell.
\XRAY Partial shell.
\DATE 31 Mar 2006

\LONGITUDE 359.1 \LATITUDE -0.5
\RAHMS 17 45 30 \DECDM -29 57
\SIZE 24 \TYPE S
\FLUX1GHZ 14 \ALPHA 0.4?
\NAMES

\RADIO Non-thermal shell in complex region, crossed by the `snake'.
\XRAY Centrally brightened.
\POINT Several compact radio sources near centre, OH masers around edge.
\DATE 19 Feb 2009

\LONGITUDE 359.1 \LATITUDE +0.9
\RAHMS 17 39 36 \DECDM -29 11
\SIZE 12\X/11 \TYPE S
\FLUX1GHZ 2? \ALPHA ?
\NAMES

\RADIO Shell, brightest in E.
\DATE 19 Feb 2009


\centerline{\hrulefill}
\label{lastpage}

\end{document}